\documentclass[12pt]{article}

\usepackage{amsmath}
\usepackage{amsfonts}
\usepackage{amssymb}
\usepackage{graphicx}
\usepackage{cite}
\usepackage{bbm}

\allowdisplaybreaks[1]

\newcommand{\imag}{\mathrm{i}}

\makeatletter
\renewcommand{\fnum@table}{\textbf{\tablename~\thetable}}
\renewcommand{\fnum@figure}{\textbf{\figurename~\thefigure}}
\makeatother

\newcounter{myenumi}

\renewcommand{\themyenumi}{\roman{myenumi}}

{\end{list}}

\newlength{\myem}
\newcounter{mysubequation}[equation]

\makeatletter
\renewcommand{\section}{\@startsection{section}{1}{0em}{-\baselineskip}%
{\baselineskip}{\normalfont\large\bfseries}}
\renewcommand{\subsection}%
{\@startsection{subsection}{2}{0em}{-0.7\baselineskip}%
{0.7\baselineskip}{\normalfont\bfseries}}
\makeatother

\newcommand{\bi}{\begin{itemize}}
\newcommand{\ei}{\end{itemize}}

\newcommand{\ldm}{\Delta m_{31}^2}
\newcommand{\sdm}{\Delta m_{21}^2}
\newcommand{\deltacp}{\delta_{\mathrm{CP}}}

\newcommand{\ie}{{\it i.e.}}

\newcommand{\eg}{{\it e.g.}}

\newcommand{\etc}{{\it etc.}}
\newcommand{\eq}{Eq.}
\newcommand{\eqs}{Eqs.}

\newcommand{\fig}{Fig.}
\newcommand{\Fig}{Fig.}
\newcommand{\figs}{Figs.}

\newcommand{\Ref}{Ref.}
\newcommand{\Refs}{Refs.}
\newcommand{\Sec}{Sec.}
\newcommand{\Secs}{Secs.}
\newcommand{\App}{Appendix}

\begin{document}

\begin{titlepage}

\renewcommand{\thefootnote}{\alph{footnote}}

\vspace*{-3.cm}
\begin{flushright}
TUM-HEP-542/04\\
\end{flushright}

\vspace*{0.5cm}

\renewcommand{\thefootnote}{\fnsymbol{footnote}}
\setcounter{footnote}{-1}

{\begin{center}
{\Large\bf Series expansions for three-flavor neutrino oscillation\\[1mm]
probabilities in matter}
\end{center}}
\renewcommand{\thefootnote}{\alph{footnote}}

\vspace*{.5cm}
{\begin{center} {\textbf{
                Evgeny~K.~Akhmedov\footnote[1]{\makebox[1.cm]{Email:}
                akhmedov@ific.uv.es},
                Robert Johansson\footnote[2]{\makebox[1.cm]{Email:}
                robert@theophys.kth.se},
                Manfred Lindner\footnote[3]{\makebox[1.cm]{Email:}
                lindner@ph.tum.de},\\[1mm]
                Tommy Ohlsson\footnote[4]{\makebox[1.cm]{Email:}
                tommy@theophys.kth.se},
                and
                Thomas Schwetz\footnote[5]{\makebox[1.cm]{Email:}
                schwetz@ph.tum.de}~
                }}
\end{center}}
\vspace*{0cm}
{\it
\begin{center}

\footnotemark[1]%
Instituto de F\'{\i}sica Corpuscular -- C.S.I.C./Universitat de 
Val\`encia, \\
Edificio Institutos de Paterna, Apt 22085, 46071 Valencia, Spain

\footnotemark[2]${}^,$\footnotemark[4]%
Division of Mathematical Physics, Department of
Physics,\\ Royal Institute of Technology (KTH) --
AlbaNova University Center,\\ Roslagstullsbacken~11,
106~91~Stockholm, Sweden 

\footnotemark[3]${}^,$\footnotemark[5]%
Theoretische Physik, Physik-Department,\\
Technische Universit\"at M\"unchen (TUM),\\
James-Franck-Stra{\ss}e, 85748~Garching bei M{\"u}nchen, Germany

\end{center}}

\vspace*{1.cm}

{\large \bf
\begin{center} Abstract \end{center}  }
We present a number of complete sets of series expansion formulas for neutrino 
oscillation probabilities in matter of constant density for three flavors.  
In particular, we study expansions in the mass hierarchy parameter $\alpha 
\equiv \Delta m_{21}^2/\Delta m_{31}^2$ and mixing parameter $s_{13} \equiv 
\sin \theta_{13}$ up to second order and expansions only in $\alpha$ and only 
in $s_{13}$ up to first order. For each type of expansion we also present the 
corresponding formulas for neutrino oscillations in vacuum.
We perform a detailed analysis of the accuracy of the different sets
of series expansion formulas and investigate which type of expansion
is most accurate in different regions of the parameter space spanned
by the neutrino energy $E$, the baseline length $L$, and the expansion
parameters $\alpha$ and $s_{13}$.
We also present the formulas for series expansions in $\alpha$ and in
$s_{13}$ up to first order for the case of arbitrary matter density
profiles. Furthermore, it is shown that in general all the 18 neutrino
and antineutrino oscillation probabilities can be expressed through
just two independent probabilities.
 
\vspace*{.5cm}

\end{titlepage}

\newpage

\renewcommand{\thefootnote}{\arabic{footnote}}
\setcounter{footnote}{0}


\section{Introduction} The discovery of neutrino oscillations in
atmospheric, solar, and reactor neutrino experiments has turned
neutrino physics into one of the most exciting and active fields of
particle physics.  By now, a significant amount of information on
neutrino properties has been obtained. The results of the atmospheric
neutrino experiments~\cite{Fukuda:1998mi,SK,Ambrosio:2003yz} and the
K2K accelerator neutrino experiment~\cite{Ahn:2002up} have allowed the
determination of the fundamental neutrino parameters $\Delta m_{31}^2$
and $\theta_{23}$ to an accuracy of about 30\% and 15\%, respectively,
while the solar neutrino experiments~\cite{solar} and the KamLAND
reactor neutrino experiment~\cite{Eguchi:2002dm} have measured the
parameters $\Delta m_{21}^2$ and $\theta_{12}$ to an accuracy of about
15\%.
In fact, this means that neutrino physics is now entering an era of
precision measurements of the neutrino oscillation parameters. Future
experiments with superbeams and neutrino factories will determine the
``atmospheric'' and ``solar'' neutrino oscillation parameters to an
accuracy of the order of 1\%.  They are also expected to measure the
elusive leptonic mixing angle $\theta_{13}$ for which at present only
an upper limit exists \cite{Apollonio:1999ae}, or to put a more
stringent limit on this angle, as well as to clarify the issues of the
neutrino mass hierarchy and possibly of leptonic CP violation.
Important information on neutrino oscillation parameters can also be
obtained from the future atmospheric, solar, reactor, and supernova
neutrino experiments.

Increasing accuracy and reach of the present and especially of
forthcoming experiments put forward new and more challenging demands
to the theoretical description of neutrino oscillations. In order to
be able to determine the fundamental neutrino parameters from the data
with high precision, one needs, among other things, very accurate
theoretical expressions for the probabilities of neutrino oscillations
in matter and in vacuum. While these probabilities can, in principle,
be calculated numerically with any requisite accuracy, it is highly
desirable to have also analytic expressions for them. Such analytic
expressions would reveal the basic dependence of the neutrino
oscillation probabilities on the fundamental neutrino parameters and
on the characteristics of the experiment and thus facilitate the
choice of the experimental setup as well as the analysis of the
data. They would also help to understand the physics underlying
various flavor transitions and to resolve the \mbox{parameter}
degeneracies and other ambiguities, such as fundamental versus
matter-induced CP violation.

The purpose of this paper is to present a collection of approximate
analytic formulas for the neutrino oscillation probabilities. We
derive a number of complete sets of series expansion formulas for the
three-flavor neutrino oscillation probabilities in matter and in
vacuum. The probabilities are expanded in the mass hierarchy parameter
$\alpha \equiv \Delta m_{21}^2/\Delta m_{31}^2$, the mixing parameter
$s_{13} \equiv \sin \theta_{13}$, or in both of them. We also study in
detail the accuracy of the obtained expressions in different regions
of the parameter space and identify the ``best choice'' in each case
of interest.

Before proceeding to present our results, we give here a brief
overview of the previous work on the subject. Analytic formulas for
three-flavor neutrino oscillation probabilities have been derived in a
number of papers.  Exact formulas for the neutrino oscillation
probabilities in vacuum can be found, \eg, in
\Ref~\cite{Bilenky:1987st}. Exact expressions can also be obtained in
the case of three-flavor neutrino oscillations in matter of constant
density~\cite{Barger:1980tf,Kim:1987vg,Zaglauer:1988gz,Ohlsson:1999xb,Xing:2000gg,Ohlsson:2001vp,Kimura:2002hb,Harrison:2003fi}.
However, the corresponding formulas are rather complicated and not
easily tractable. This also applies to the exact analytic three-flavor
formulas obtained for some special cases of non-uniform matter
density: linear matter density \cite{Lehmann:2000ey} and exponentially
varying matter density \cite{Osland:1999et}.

Approximate analytic formulas for three-flavor neutrino oscillation
probabilities in matter have been derived in a number of papers, see
\eg,
\Refs~\cite{Kuo:1986sk,Smirnov:1987mk,Kuo:1987zx,Petcov:1987qg,Ohlsson:2001et,D'Olivo:1996nk}. In
\Refs~\cite{Kuo:1986sk,Smirnov:1987mk}, the
Mikheyev--Smirnov--Wolfenstein (MSW) resonances~\cite{MSW} in matter
of varying density were studied assuming large separation and
independence of the high-density and low-density resonances.  In
\Refs~\cite{Smirnov:1987mk,Kuo:1987zx,Ohlsson:2001et}, the adiabatic
approximation was used to derive three-flavor neutrino oscillation
probabilities in matter of varying density, whereas in
\Ref~\cite{D'Olivo:1996nk}, the Magnus expansion for the time
evolution operator was used, which is equivalent to the anti-adiabatic
approximation.

In a number of papers an approach similar to ours was adopted, \ie,
the neutrino oscillation probabilities were expanded in the small
parameters $\alpha$, $s_{13}$, or in both of them.  In
\Ref~\cite{Akhmedov:1998xq}, exact analytic expressions for
three-flavor neutrino oscillation probabilities in matter with an
arbitrary density profile were obtained in the limit $\alpha \to 0$ by
reducing the problem to an effective two-flavor one. In
\Ref~\cite{Peres:1999yi}, a similar approach was employed to obtain
the neutrino oscillation probabilities in matter of arbitrarily
varying density in the limit $\theta_{13} \to 0$, whereas in
\Refs~\cite{Akhmedov:2001kd,Peres:2002ri}, expressions up to first
order in $s_{13}$ were derived. For the case of matter of constant
density, the limit $\alpha \to 0$ was considered in
\Ref~\cite{Yasuda:1999uv}. Expansions up to first order in $\alpha$
were carried out in \Refs~\cite{Freund:1999gy,Mocioiu:2001jy}. In
\Refs~\cite{Arafune:1997hd,Brahmachari:2003bk}, both the solar mass
squared difference $\sdm$ and matter effects were treated as
perturbations and the transition probabilities up to first order in
them were derived. Expansions in both $\alpha$ and $s_{13}$ were used
in
\Refs~\cite{Cervera:2000kp,Freund:2001pn,Freund:2001ui,Barger:2001yr}.
We note that the neutrino oscillation probabilities derived in most of
the above-mentioned papers either did not constitute a complete set
(\ie, a set of probabilities from which the probabilities in all
channels can be obtained), or contained expressions from which some
terms were missing, especially in the case of the probabilities
$P(\nu_\mu\to \nu_\mu)$ and $P(\nu_\mu\to \nu_\tau)$. Thus, to the
best of our knowledge, our study is the first one in which, for the
case of matter of constant density, complete and consistent expansions
in $\alpha$ and $s_{13}$ to second order and expansions only in
$\alpha$ and only in $s_{13}$ up to first order are performed.

The paper is organized as follows. In \Sec~\ref{sec:notation} we set
the general formalism and notation, discuss some general relations
satisfied by the oscillation probabilities, and show that all the 18
neutrino and antineutrino oscillation probabilities can be expressed
through just two independent probabilities.  In \Sec~\ref{sec:double}
we present a series expansion of the neutrino oscillation
probabilities up to second order in both $\alpha$ and $s_{13}$ (the
so-called ``double expansion''), whereas in \Secs~\ref{sec:alpha} and
\ref{sec:s13} we consider the probabilities expanded up to first order
in $\alpha$ and $s_{13}$, respectively (``single expansions'').  In
all three cases, we give the probabilities for matter of constant
density (\Secs~\ref{sec:double_const}, \ref{sec:alpha_const}, and
\ref{sec:s13_const}) and in vacuum (\Secs~\ref{sec:double_vac},
\ref{sec:alpha_vac}, and \ref{sec:s13_vac}). 
We also compare our formulas with the corresponding expressions existing in 
the literature (when available), pointing out agreements and disagreements. 
In \Sec~\ref{sec:quality}, we discuss the qualitative behavior of the
neutrino oscillation probabilities, the relevance of matter effects,
and give a detailed evaluation of the accuracy of the various
formulas. Furthermore, we comment on the application of our
probability formulas to neutrino oscillation experiments.  We
summarize our results in \Sec~\ref{sec:conclusions}. Finally, several
methods that have been used to derive the formulas are presented in
the appendices.  In \App~\ref{app:diagonalization}, we describe the
perturbative diagonalization of the effective Hamiltonian of the
neutrino system in the case of matter of constant density, while in
\App~\ref{app:pert_evol}, the details of the perturbative expansion of
the neutrino evolution equation in the case of arbitrary matter
density profiles are given.

\section{General formalism and notation}
\label{sec:notation} 

We consider three-flavor neutrino oscillations and adopt the standard 
parameterization of the leptonic mixing matrix $U$ \cite{PDG}: 
\begin{align}
U &=  O_{23} U_\delta O_{13} U^\dagger_\delta O_{12} \nonumber\\
  &= \left( \begin{matrix} c_{12} c_{13} &
    s_{12} c_{13} & s_{13} {\rm e}^{-{\rm i} \delta_{\rm CP}} \\ -s_{12}
    c_{23} - c_{12} s_{13} s_{23} {\rm e}^{{\rm i} \delta_{\rm CP}} & c_{12}
    c_{23} - s_{12} s_{13} s_{23} {\rm e}^{{\rm i} \delta_{\rm CP}} & c_{13}
    s_{23} \\ s_{12} s_{23} - c_{12} s_{13} c_{23} {\rm e}^{{\rm i}
      \delta_{\rm CP}} & -c_{12} s_{23} - s_{12} s_{13} c_{23} {\rm
      e}^{{\rm i} \delta_{\rm CP}} & c_{13} c_{23} \end{matrix} \right) \,. 
   \label{eq:U}   
\end{align}
Here $O_{ij}$ is the orthogonal rotation matrix in the $ij$-plane
which depends on the mixing angle $\theta_{ij}$,
$U_\delta=\mathrm{diag} (1,1,{\rm e}^{{\rm i}\delta_{\rm CP}})$,
$\delta_{\rm CP}$ being the Dirac-type CP-violating phase, $s_{ij}
\equiv \sin \theta_{ij}$ and $c_{ij} \equiv \cos \theta_{ij}$. In the
three-flavor case there are also, in general, two Majorana-type
CP-violating phases; however, these phases do not affect neutrino
oscillations, and therefore will not be considered here. Without loss
of generality, one can assume all the mixing angles to lie in the
first quadrant (\ie, between 0 and $\pi/2$), while the CP-violating
phase $\delta_{\rm CP}$ is allowed to lie in the interval $[0,
\,2\pi]$.

Let us denote by $P_{\alpha\beta} \equiv P(\nu_\alpha\to\nu_\beta)$
the probability of transition from a neutrino flavor $\alpha$ to a
neutrino flavor $\beta$, and similarly for antineutrino flavors, \ie,
$P_{\bar\alpha \bar\beta} \equiv
P(\bar\nu_\alpha\to\bar\nu_\beta)$. In general, the three-flavor
neutrino oscillation probabilities in matter $P_{\alpha\beta}$ depend
on eight parameters and one function:
\begin{equation}
P_{\alpha\beta} = P_{\alpha\beta}(\Delta m_{21}^2, \Delta m_{31}^2,
\theta_{12}, \theta_{13}, \theta_{23}, \delta_{\rm CP}; E, L,
V(x)), \quad \alpha,\beta = e,\mu,\tau \,.
\label{eq:Pvars}
\end{equation}
Here $\Delta m_{ij}^2 \equiv m^2_i - m^2_j$ are the neutrino mass
squared differences, $E$ is the neutrino energy, $L$ is the baseline
length, and $V(x)$ is the matter-induced effective potential, $x \in
[0,L]$ being the coordinate along the neutrino path. The neutrino mass
squared differences, the leptonic mixing angles, and the CP-violating
phase are fundamental parameters and thus experiment-independent,
whereas the neutrino energy, the baseline length, and the
matter-induced effective potential vary from experiment to experiment.
The present best-fit values and 3$\sigma$ allowed ranges of the
fundamental neutrino parameters found in a recent global fit of the
neutrino oscillation data \cite{Maltoni:2003da} are summarized in
Table~\ref{tab:parameters}.  Unless otherwise stated, all calculations
in the present paper are performed for the following values of the
neutrino parameters: $\Delta m_{21}^2 = 7 \cdot 10^{-5} \,{\rm eV}^2$,
$\theta_{12} = 33^\circ$, and $\theta_{23} =45^\circ$. For the
atmospheric mass squared difference $\ldm$ we adopt the current
best-fit value given by the Super-Kamiokande Collaboration, $|\Delta
m_{31}^2| = 2 \cdot 10^{-3} \, {\rm eV}^2$ \cite{SK}, which is
slightly smaller than the value given in Table~\ref{tab:parameters}.
The sign of $\ldm$ is related to the neutrino mass hierarchy: for the
normal (inverted) hierarchy one has $\ldm > 0$ ($\ldm < 0$). For the
leptonic mixing angle $\theta_{13}$ we allow values below the 90\%
C.L. upper bound found in the global fit of the neutrino oscillation
data \cite{Maltoni:2003da} for $|\ldm|$ fixed at $2 \cdot 10^{-3} \,
{\rm eV}^2$:
\begin{equation}
\theta_{13} \lesssim 10.8^\circ\,, \quad\mbox{or}\quad 
s_{13} \equiv \sin \theta_{13} \lesssim 0.19 \,, \quad\mbox{or}\quad 
\sin^2 2\theta_{13} \lesssim 0.14 \,.
\label{eq:th13bound}
\end{equation}
For the CP violation phase $\deltacp$ we allow values between 
$0$ and $2\pi$.

\begin{table}[t!]
\begin{center}
\begin{tabular}{ccc}
\hline
\\[-4mm]
Parameter & Best-fit value & Range ($3\sigma$)\\
\\[-4mm]
\hline
\\[-4mm]
$\Delta m_{21}^2$ & $6.9 \cdot 10^{-5} \, {\rm eV}^2$ & $(5.4 \div
9.5) \cdot 10^{-5} \, {\rm eV}^2$ \\
$|\Delta m_{31}^2|$ & $2.6 \cdot 10^{-3} \, {\rm eV}^2$ & $(1.5 \div
3.7) \cdot 10^{-3} \, {\rm eV}^2$ \\
$\theta_{12}$ & $33.2^\circ$ & $28.6^\circ \div 38.6^\circ$ \\
$\theta_{13}$ & $4.4^\circ$ & $0 \div 13.4^\circ$ \\
$\theta_{23}$ & $46.1^\circ$ & $33.8^\circ \div 58.1^\circ$ \\
$\delta_{\rm CP}$ & - & $0 \div 2\pi$\\
\\[-4mm]
\hline
\end{tabular}
\caption{\label{tab:parameters} Present best-fit values and $3\sigma$
  allowed ranges of the fundamental neutrino parameters from a three-flavor
  fit to global neutrino oscillation data~\protect\cite{Maltoni:2003da}.}
\end{center}
\vspace*{-2mm}
\end{table}

Inspecting the values of the fundamental neutrino parameters in
Table~\ref{tab:parameters}, one can identify two natural candidates for 
small expansion parameters of the neutrino oscillation probabilities. 
These are the small leptonic mixing angle $\theta_{13}$ (or, equivalently, 
$s_{13}$) and the mass hierarchy parameter
\begin{equation}
\alpha \equiv \frac{\Delta m_{21}^2}{\Delta m_{31}^2} \simeq 0.026 
\,,\quad 
 (0.018)~0.021 \lesssim \alpha \lesssim 0.036~(0.053) 
\quad\mbox{at 90\% C.L. ($3\sigma$)\,,}
\label{eq:def_alpha}
\end{equation}
where we have taken the current best-fit value and allowed ranges from
\Ref~\cite{Maltoni:2003da}. In the following, we will derive a number
of formulas for series expansions of the neutrino oscillation
probabilities in these small quantities. Comparing
\eqs~(\ref{eq:th13bound}) and (\ref{eq:def_alpha}), one realizes that
current data constrain the parameter $\alpha$ to a relatively narrow range,
while $s_{13}$ is only bounded from above and might be significantly 
larger or smaller than $\alpha$.  The relative size of these two 
expansion parameters will be important for the validity of a given type of
expansion.

In order to find the neutrino oscillation probabilities for a given 
experimental setup, one has, in general, to solve the Schr{\"o}dinger equation 
for the neutrino vector of state in the flavor basis $|\nu(t)\rangle =
(\begin{matrix} \nu_e(t) & \nu_\mu(t) & \nu_\tau(t) \end{matrix})^T$: 
\begin{equation}\label{eq:schrodinger}
{\rm i} \frac{{\rm d}}{{\rm d} t} |\nu(t) \rangle = H |\nu(t) \rangle
\end{equation}
with the effective Hamiltonian
\begin{equation}\label{eq:ham1}
H \simeq \frac{1}{2E} U \,\mathrm{diag}(0 , \sdm , \ldm) U^\dagger + 
\mathrm{diag}(V,0,0) \,.
\end{equation}
Here $V$ is the charged-current contribution to the matter-induced
effective potential of $\nu_e$~\cite{MSW}. We have disregarded the
neutral-current contributions to the neutrino potentials in matter,
since they are the same for $\nu_e$, $\nu_\mu$, and
$\nu_\tau$~\footnote{Up to tiny radiative corrections
\cite{Botella:1987wy} which are negligible except at very high
densities available, \eg, inside supernovae.}  and so do not affect
neutrino oscillations.
Note that \eq~(\ref{eq:ham1}) holds for neutrinos, whereas for antineutrinos 
one has to perform the replacements
\begin{equation}
U \to U^* \:,\quad V \to -V \,.
\label{eq:antinu}
\end{equation}
The potential $V(x)$ is given in convenient units by
\begin{equation}
V(x) \simeq 7.56 \times 10^{-14} 
\left( \frac{\rho(x)}{\mathrm{g/cm^3}} \right) Y_e(x) \;\; 
\mathrm{eV} \,,
\end{equation}
where $\rho(x)$ is the matter density along the neutrino path and $Y_e(x)$ 
is the number of electrons per nucleon. For the matter of the Earth one has, 
to a very good accuracy, $Y_e\simeq 0.5$. 

For many practical applications (such as long-baseline accelerator
experiments, as well as oscillations of atmospheric, solar, and 
supernova neutrinos inside the Earth when they do not cross the
Earth's core) it is a very good approximation to assume that the matter
density along the neutrino trajectory is constant (see, \eg,
\Refs~\cite{Nicolaidis:1988fe,Liu:1998nb, Freund:1999vc}). Typical values
for the matter density are $\rho_{\rm crust} \simeq 3 \, {\rm g/cm^3}$
in the Earth's crust and $\rho_{\rm mantle} \simeq 4.5 \, {\rm
g/cm^3}$ in its mantle. In situations where the neutrinos also cross
the Earth's core or for strongly varying matter density profiles
like those inside the Sun or supernovae, the constant matter density
approximation is not valid.

The neutrino oscillation probabilities can be found as $P_{\alpha\beta} =
|S_{\beta\alpha}(t,t_0)|^2$, where $S(t,t_0)$ is the evolution matrix
such that 
\begin{equation}
|\nu(t)\rangle=S(t, t_0) |\nu(t_0)\rangle\,,\qquad\qquad
S(t_0, t_0)=\mathbbm{1}\,.
\label{eq:S}
\end{equation}
Note that $S(t,t_0)$ satisfies the same Schr{\"o}dinger equation,
\eq~(\ref{eq:schrodinger}), as $|\nu(t)\rangle$.  In the case of
matter of constant density, the evolution matrix can be obtained by
diagonalizing the Hamiltonian in \eq~(\ref{eq:ham1}) according to 
$H = U' \hat H U'^\dagger$, where $U'$ is the leptonic mixing matrix in 
matter and $\hat H = \mathrm{diag}(E_1,E_2,E_3)$. The evolution matrix is 
then given by
\begin{equation}\label{eq:evolution}
S_{\beta\alpha} (t,t_0) = 
\sum_{i=1}^3 (U'_{\alpha i})^* U'_{\beta i} {\rm e}^{-{\rm i} E_i L}
\,, \qquad \alpha,\beta = e,\mu,\tau \,,
\end{equation}
where we have identified $L \equiv t - t_0$.

Before presenting our results for the neutrino oscillation
probabilities, we discuss some of their general properties as well as
relations between them. First, we note that \eq~(\ref{eq:antinu})
implies that one can relate the oscillation probabilities for
antineutrinos to those for neutrinos by
\begin{equation}\label{eq:Panti}
P_{\bar{\alpha}\bar{\beta}} = P_{\alpha\beta}(\delta_{\rm CP} \to -
\delta_{\rm CP}, \, V \to -V) \,, \qquad \alpha,\beta = e,\mu,\tau \,.
\end{equation}
Second, in general, \ie, both in vacuum and in matter with an arbitrary 
density profile, it follows from the unitarity of $S(t,t_0)$ (conservation of 
probability) that
\begin{equation}
\sum_{\alpha} P_{\alpha\beta} = \sum_{\beta}
P_{\alpha\beta} = 1 \,, \qquad \alpha,\beta = e,\mu,\tau \,.
\label{eq:probsum}
\end{equation}
These relations imply that five out of the nine neutrino oscillation
probabilities can be expressed in terms of the other
four \cite{deGouvea:2000un}.
 
Besides these general properties, there exists an additional symmetry
due to the specific parameterization of the mixing matrix given in
\eq~(\ref{eq:U}) and the fact that the rotation matrix $O_{23}$ commutes 
with the matter potential term of the Hamiltonian in \eq~(\ref{eq:ham1}).
It is easy to show that the evolution matrix can be written as
\begin{equation}
S(t, t_0) = O_{23} S'(t, t_0) O_{23}^T
\label{eq:SS'}
\end{equation}
where $S'(t,t_0)$ does not depend on $\theta_{23}$. This can be used
to prove some useful relations between the probabilities.  Let us
denote~\footnote{The transformation in \eq~(\ref{eq:notat}) can be
achieved, \eg, through the substitution $\theta_{23}\to
\theta_{23}+\pi/2$ or $\theta_{23}\to \theta_{23}+3\pi/2$.}
\begin{equation}
\tilde{P}_{\alpha\beta}\equiv P_{\alpha\beta}(
s_{23}^2\leftrightarrow c_{23}^2\,,\; 
\sin 2\theta_{23}\to -\sin 2\theta_{23})\,, \qquad \alpha,\beta =
e,\mu,\tau \,.
\label{eq:notat}
\end{equation}
Using \eqs~(\ref{eq:SS'}) and (\ref{eq:notat}), one can 
readily show that 
\begin{equation}
P_{e\tau}=\tilde{P}_{e\mu}\,,\qquad\quad
P_{\tau\mu}=\tilde{P}_{\mu\tau}\,, \qquad\quad
P_{\tau\tau}=\tilde{P}_{\mu\mu}\,, 
\label{eq:mutau-sym}
\end{equation}
while $P_{ee}$ turns out to be independent of $\theta_{23}$.

Out of the three conditions in \eq~(\ref{eq:mutau-sym}), only two are
independent, as each of them can be derived from the other two and the
unitarity conditions (\ref{eq:probsum}). Hence, the number of
independent neutrino oscillation probabilities is reduced to
two. Thus, we come to the important conclusion that all the nine
neutrino oscillation probabilities can be expressed through just two
independent probabilities provided that their dependence on the mixing
angle $\theta_{23}$ is known. However, the choice of these independent
probabilities is restricted: they should not include $P_{ee}$, which
is independent of $\theta_{23}$; nor should they be a pair of the
probabilities, which are T-reverse of each other, or go into each
other (or T-reverse of each other) under the transformation
$s_{23}^2\leftrightarrow c_{23}^2\,,\;\sin 2\theta_{23}\to -\sin
2\theta_{23}$.

One possible choice, and the one that we will use, is $P_{e\mu}$ and
$P_{\mu\tau}$. For completeness, we give here the expressions for all
the other neutrino oscillation probabilities in terms of these two.
Using \eqs~(\ref{eq:probsum}) and (\ref{eq:mutau-sym}), one easily
finds
\begin{align}
\label{eq:Pee_gen}
P_{ee} &= 1-(P_{e\mu}+\tilde{P}_{e\mu})\,, \\
P_{e\tau} &= \tilde{P}_{e\mu} \,, \\
P_{\mu e} &= P_{e\mu}-P_{\mu\tau}+\tilde{P}_{\mu\tau}\,, \\
P_{\mu\mu} &= 1-P_{e\mu}-\tilde{P}_{\mu\tau}\,, \\
P_{\tau e} &= \tilde{P}_{e\mu}+P_{\mu\tau}-\tilde{P}_{\mu\tau}\,, \\
P_{\tau\mu} &= \tilde{P}_{\mu\tau}\,, \\
P_{\tau\tau} &= 1-(\tilde{P}_{e\mu}+P_{\mu\tau})\,. 
\label{eq:relations}
\end{align}

In addition to the above relations, one can study the transformations of the 
neutrino oscillation probabilities under the time reversal $P_{\alpha\beta}\to 
P_{\beta\alpha}$. It can be shown \cite{Akhmedov:2001kd} that in matter 
with an arbitrary density profile 
\begin{equation}
P_{\beta\alpha} = 
P_{\alpha\beta}(\delta_{\rm CP} \to - \delta_{\rm CP}, \, V(x)\to 
V_{\rm rev}(x))\,, \qquad \alpha,\beta = e,\mu,\tau \,,
\label{eq:Trev1}
\end{equation}
where $V_{\rm rev}(x)$ is the ``reverse'' potential, which corresponds
to the interchanged positions of the neutrino source and the detector. In the 
case of symmetric matter density profiles (including matter of constant 
density), $V_{\rm rev}(x)=V(x)$, and \eq~(\ref{eq:Trev1}) simplifies to
\begin{equation}
P_{\beta\alpha} = 
P_{\alpha\beta}(\delta_{\rm CP} \to - \delta_{\rm CP}) \,, \qquad
\alpha,\beta = e,\mu,\tau \,.
\label{eq:T}
\end{equation}
While \eq~(\ref{eq:T}) does not further reduce the number of independent 
probabilities, it yields relations which can be useful for cross-checking the 
formulas for $P_{\alpha\beta}$ in the case of matter with symmetric density 
profiles.  
 
By applying the rule given in \eq~(\ref{eq:Panti}), one can obtain
from \eqs~(\ref{eq:Pee_gen})--(\ref{eq:relations}) the corresponding
probabilities for the antineutrino oscillations. Thus, the expressions
for all 18 probabilities of neutrino and antineutrino oscillations can
be found from the formulas for just two independent neutrino
oscillation probabilities, which, as was already mentioned, we choose
to be $P_{e\mu}$ and $P_{\mu\tau}$.  In order to be more explicit, we
will in some cases also give formulas for additional neutrino
oscillation channels.

In the following \Secs~\ref{sec:double}, \ref{sec:alpha}, and
\ref{sec:s13}, we give our results for the various series expansions
of the neutrino oscillation probabilities.  We will adopt the
following abbreviations:
\begin{align}
\Delta &\equiv \frac{\Delta m^2_{31} L}{4 E} \,, 
\label{eq:def_Delta}\\
A &\equiv \frac{2EV}{\Delta m^2_{31}} = \frac{VL}{2\Delta}\,.
\label{eq:def_A}
\end{align}

\section{Series expansion up to second order in
$\mbox{\boldmath$\alpha$}$ and $\mbox{\boldmath$s_{13}$}$}
\label{sec:double}

\subsection{Matter of constant density }
\label{sec:double_const}

In this section, we present the series expansion formulas for three-flavor
neutrino oscillation probabilities in matter of constant density up to second
order in both $\alpha$ and $s_{13}$. The probabilities are calculated by 
diagonalizing the Hamiltonian~(\ref{eq:ham1}) up to second order in these  
parameters, as described in \App~\ref{app:perturbation}. We find for the 
eigenvalues of the Hamiltonian
\begin{align}
E_1 &\simeq \frac{\Delta m_{31}^2}{2E} \,
     \left( A + \alpha \, s_{12}^2 + s_{13}^2 \, \frac{A}{A-1} + 
     \alpha^2 \,\frac{\sin^22\theta_{12}}{4 A} \right) \,, \label{eq:EV1}\\
E_2 &\simeq \frac{\Delta m_{31}^2}{2E} \,
     \left( \alpha \, c_{12}^2  - \alpha^2 \, 
     \frac{\sin^22\theta_{12}}{4 A} \right) \,, \label{eq:EV2}\\
E_3 &\simeq \frac{\Delta m_{31}^2}{2E} \,
     \left( 1 - s_{13}^2 \, \frac{A}{A - 1}\right) \,. \label{eq:EV3}
\end{align}
Calculating the mixing matrix in matter $U'$ as described in
\App~\ref{app:perturbation} and using \eq~(\ref{eq:evolution}) for the
evolution matrix $S$, it is straightforward (although somewhat
tedious) to derive the following expressions for the neutrino oscillation
probabilities:
\begin{align}
P_{ee} &= 1 - \alpha^2 \, \sin^2 2\theta_{12} \, \frac{\sin^2
  A\Delta}{A^2} - 4 \, s_{13}^2 \, \frac{\sin^2 (A-1)\Delta}{(A-1)^2} \,,
\label{eq:Pee} \\[3mm]
P_{e\mu} &= \alpha^2 \, \sin^2 2\theta_{12} \, c_{23}^2 \frac{\sin^2
         A\Delta}{A^2} + 4 \, s_{13}^2 \, s_{23}^2 \frac{\sin^2
         (A-1)\Delta}{(A-1)^2} \nonumber\\
         &+ 2 \, \alpha \, s_{13} \, \sin 2\theta_{12} \, 
         \sin2\theta_{23} \cos(\Delta - \delta_{\rm CP}) \, \frac{\sin
         A\Delta}{A} \, \frac{\sin (A-1)\Delta}{A-1} \,,
\label{eq:Pem}\\[3mm]
P_{e\tau} &= \alpha^2 \, \sin^2 2\theta_{12} \, s_{23}^2 \frac{\sin^2
         A\Delta}{A^2} + 4 \, s_{13}^2 \, c_{23}^2 \frac{\sin^2
         (A-1)\Delta}{(A-1)^2} \nonumber\\
         &- 2 \, \alpha \, s_{13} \, \sin 2\theta_{12} \, 
         \sin2\theta_{23} \cos(\Delta - \delta_{\rm CP}) \, \frac{\sin
         A\Delta}{A} \, \frac{\sin (A-1)\Delta}{A-1} \,,
\label{eq:Pet}\\[3mm]
P_{\mu\mu} &= 1 - \sin^2 2\theta_{23} \, \sin^2\Delta +
   \alpha \, c_{12}^2 \, \sin^2 2\theta_{23} \, \Delta 
   \, \sin 2 \Delta  \nonumber\\
   &- \alpha^2 \, \sin^2 2\theta_{12} \, c_{23}^2 \, \frac{\sin^2
      A\Delta}{A^2} - \alpha^2 \, c_{12}^4 \, \sin^2 2\theta_{23} \,
   \Delta^2 \, \cos 2 \Delta \nonumber\\
   &+ \frac{1}{2A} \, \alpha^2 \, \sin^2 2\theta_{12} \,
      \sin^22\theta_{23} \left( 
       \sin\Delta \, \frac{\sin A\Delta}{A} \, \cos (A-1)\Delta - 
       \frac{\Delta}{2} \, \sin2\Delta \right) \nonumber\\
   &- 4 \, s_{13}^2 \, s_{23}^2  \frac{\sin^2 (A-1)\Delta}{(A-1)^2}
   \nonumber\\
   &- \frac{2}{A-1} \, s_{13}^2 \, \sin^22\theta_{23} \left(
       \sin\Delta \, \cos A\Delta \, \frac{\sin (A-1)\Delta}{A-1} - 
       \frac{A}{2} \Delta \, \sin2\Delta \right) \nonumber\\
   &- 2 \, \alpha \, s_{13} \, \sin 2\theta_{12} \,  \sin 2\theta_{23}
       \, \cos\delta_{\rm CP} \, \cos\Delta \, \frac{\sin A\Delta}{A} \,
   \frac{\sin (A-1)\Delta}{A-1} \nonumber\\
   &+ \frac{2}{A-1} \, \alpha \, s_{13} \, \sin 2\theta_{12} \, 
       \sin2\theta_{23} \, \cos2\theta_{23} \, \cos\delta_{\rm CP} \,
   \sin\Delta  
   \left( A \sin\Delta - \frac{\sin A\Delta}{A} \, \cos (A-1)\Delta \right) \,,
\label{eq:Pmm}\\[3mm]
P_{\mu\tau} &= \sin^2 2\theta_{23} \, \sin^2\Delta -
       \alpha \, c_{12}^2 \, \sin^2 2\theta_{23} \, \Delta 
       \, \sin 2 \Delta + \alpha^2 \, c_{12}^4 \, \sin^2 2\theta_{23} \,
       \Delta^2 \, \cos 2 \Delta \nonumber\\
   &- \frac{1}{2A} \, \alpha^2 \, \sin^2 2\theta_{12} \,
      \sin^2 2\theta_{23} \left(
       \sin\Delta \, \frac{\sin A\Delta}{A} \, \cos (A-1)\Delta - 
       \frac{\Delta}{2} \, \sin2\Delta \right) \nonumber\\
   &+ \frac{2}{A-1} \, s_{13}^2 \, \sin^22\theta_{23} \left(
       \sin\Delta \, \cos A\Delta \, \frac{\sin (A-1)\Delta}{A-1} - 
       \frac{A}{2} \Delta \, \sin2\Delta \right) \nonumber\\
   &+ 2 \, \alpha \, s_{13} \, \sin 2\theta_{12} \,  \sin2\theta_{23}
       \, \sin\delta_{\rm CP} \, \sin\Delta \, \frac{\sin A\Delta}{A}
       \, \frac{\sin (A-1)\Delta}{A-1} \nonumber\\
   &- \frac{2}{A-1} \, \alpha \, s_{13} \, \sin 2\theta_{12} \, 
       \sin2\theta_{23} \, \cos2\theta_{23} \, \cos\delta_{\rm CP} \,
       \sin\Delta
       \left( A \sin\Delta - \frac{\sin A\Delta}{A} \, 
       \cos (A-1)\Delta \right) \,.
\label{eq:Pmt} 
\end{align}

Formally, our calculations are based upon the approximations $\alpha,s_{13} 
\ll 1$ and no explicit assumptions about the values of $L/E$ are made. 
However, we remark that the series expansion formulas 
(\ref{eq:Pee})--(\ref{eq:Pmt}) are no longer valid as soon as $\alpha\Delta 
= \Delta m^2_{21} L/(4E)$ becomes of order unity, \ie, when the oscillatory 
behavior due to the ``solar'' mass squared difference $\sdm$ becomes 
relevant. This can happen for very long baselines and/or very low energies. 
See also \Sec~\ref{sec:quality} for a detailed discussion of the accuracy of 
these formulas.

From \eqs~(\ref{eq:EV1})--(\ref{eq:EV3}) one can see that in vacuum 
($A=0$) and at the atmospheric resonance ($A=1$) the expressions for the 
eigenvalues are divergent, and one would expect the expansion to break 
down. In these cases, two out of the three eigenvalues of the unperturbed 
Hamiltonian are degenerate and, strictly speaking, the degenerate perturbation 
theory rather than the ordinary one should be employed. However, it turns out 
that, though the eigenvalues (\ref{eq:EV1})--(\ref{eq:EV3}) are divergent, the 
neutrino oscillation probabilities are finite in the limits $A \to 0$ and $A 
\to 1$. The reason for this interesting behavior is a cancellation of 
divergences between the eigenvalues and the matrix elements of the leptonic 
mixing matrix in the calculation of the evolution matrix according to 
\eq~(\ref{eq:evolution}). 
In particular, in the limit $A\to 0$, the correct vacuum neutrino
oscillation probabilities are obtained.

We shall now compare the above results with those existing in the
literature. Equations~(\ref{eq:EV1})--(\ref{eq:Pet}) have previously
been derived in \Ref~\cite{Cervera:2000kp} and confirmed in
\Ref~\cite{Freund:2001pn}. Expressions (\ref{eq:Pmm}) and
(\ref{eq:Pmt}) are new.  In \Ref~\cite{Barger:2001yr}, an expression
for $P_{\mu\tau}$ was found to first order in $\alpha$, including the
${\cal O}(\alpha \, s_{13})$ term, which can be compared with the
corresponding terms in our \eq~(\ref{eq:Pmt}). We find that, while our
${\cal O}(1)$ and ${\cal O}(\alpha)$ terms coincide with those in
\eq~(A3) of \Ref~\cite{Barger:2001yr}, our ${\cal O}(\alpha \, s_{13})$
term is quite different. In particular, we disagree with the statement
in \Ref~\cite{Barger:2001yr} that to order $\alpha$ the probability
$P_{\mu\tau}$ does not depend on the CP-violating phase $\delta_{\rm
CP}$. The existence of a term proportional to $\alpha \sin\delta_{\rm
CP}$ in $P_{\mu\tau}$ is actually expected, since in matter of
constant density the phase $\delta_{\rm CP}$ is the sole source of
T-violation, and up to the sign, the T-odd terms in all three
transition probabilities must be the same \cite{Akhmedov:1998xq}. This
is indeed seen in \eqs~(\ref{eq:Pem}), (\ref{eq:Pet}), and
(\ref{eq:Pmt}).  We also note that the term of order $\alpha \, s_{13}$
in \eq~(A3) of \Ref~\cite{Barger:2001yr} diverges at the atmospheric
resonance ($A=1$), while our expression (\ref{eq:Pmt}) is regular at
all physical values of parameters.

\subsection{Vacuum neutrino oscillation probabilities up to second order in 
$\mbox{\boldmath$\alpha$}$ and $\mbox{\boldmath$s_{13}$}$}
\label{sec:double_vac}

The vacuum neutrino oscillation probabilities up to second order in
$\alpha$ and $s_{13}$ are given by
\begin{align}
P_{ee}^\mathrm{vac}
  &= 1 - \alpha^2 \, \sin^2 2\theta_{12} \, \Delta^2 - 4 \, s_{13}^2 \,
  \sin^2 \Delta \,, \\[1mm]
P_{e\mu}^\mathrm{vac} 
  &= \alpha^2 \, \sin^2 2\theta_{12} \, c_{23}^2 \,
  \Delta^2 + 4 \, s_{13}^2 \, s_{23}^2 \, \sin^2\Delta 
  + 2 \, \alpha \, s_{13} \, \sin2\theta_{12} \sin2\theta_{23} \,
  \cos(\Delta-\delta_{\rm CP}) \, \Delta \, \sin\Delta \,, \\[1mm]
P_{e \tau}^\mathrm{vac} &= \alpha^2 \, \sin^2 2\theta_{12} \, s_{23}^2
\, \Delta^2 + 4 \, s_{13}^2 \, c_{23}^2 \, \sin^2\Delta - 2 \, \alpha
\, s_{13} \, \sin2\theta_{12} \sin2\theta_{23} \,
\cos(\Delta-\delta_{\rm CP}) \, \Delta \, \sin\Delta \,, \\[3mm]
P_{\mu \mu}^\mathrm{vac} &= 1 - \sin^2 2\theta_{23} \, \sin^2 \Delta +
  \alpha \, c_{12}^2 \, \sin^2 2\theta_{23} \, \Delta \, \sin 2\Delta
  \nonumber\\ &- \alpha^2 \, \Delta^2 \left[ \sin^2 2\theta_{12} \,
  c_{23}^2 + c_{12}^2 \, \sin^2 2\theta_{23} \left(\cos 2\Delta 
  - s^2_{12} \right) \right] + 4 \, s_{13}^2 \, s_{23}^2 \, \cos
  2\theta_{23} \, \sin^2 \Delta \nonumber\\ &- 2 \, \alpha \, s_{13}
  \, \sin 2\theta_{12} \, s_{23}^2 \, \sin 2\theta_{23} \, \cos
  \delta_{\rm CP} \, \Delta \, \sin 2\Delta\,, \\[3mm]
P_{\mu \tau}^\mathrm{vac} &= \sin^2 2\theta_{23} \, \sin^2 \Delta -
  \alpha \, c_{12}^2 \, \sin^2 2\theta_{23} \, \Delta \, \sin 2\Delta
  \nonumber\\ &+ \alpha^2 \, \sin^2 2\theta_{23} \, \Delta^2 \left(
  c_{12}^4 \, \cos 2\Delta - \frac{1}{2} \, \sin^2 2\theta_{12} \,
  \sin^2 \Delta \right) \nonumber\\ &- 2 \, s_{13}^2 \, \sin^2
  2\theta_{23} \, \sin^2 \Delta \nonumber\\ &+ 2 \, \alpha \, s_{13}
  \, \sin 2\theta_{12} \sin 2\theta_{23} \, (\sin \delta_{\rm CP} \,
  \sin \Delta - \cos 2\theta_{23} \, \cos \delta_{\rm CP} \, \cos
  \Delta) \, \Delta \, \sin \Delta \,.
\end{align}

\section{Series expansion up to first order in $\mbox{\boldmath$\alpha$}$}
\label{sec:alpha}

In this section, we expand the probabilities up to first order in the small 
parameter $\alpha$ while keeping their exact dependence on $\theta_{13}$. 
One expects these formulas to be useful for relatively large values of 
$s_{13}$. In addition, they will correctly account for the ``atmospheric'' 
resonance driven by the parameters $\Delta m_{31}^2$ and $\theta_{13}$. 

\subsection{Matter of constant density}
\label{sec:alpha_const}

The eigenvalues of the Hamiltonian (\ref{eq:ham1}) up to first order in 
$\alpha$ are 
\begin{align}
E_1 &\simeq 
    \frac{\Delta m_{31}^2}{2E} \, \alpha \, c_{12}^2 \,,
    \label{eq:alpha_EV_1} \\
E_2 &\simeq \frac{\Delta m_{31}^2}{2E} \left[
    \frac{1}{2}\left(1+A-C_{13}\right) + \frac{1}{2C_{13}} \alpha \, s_{12}^2
    \left( C_{13} + 1 - A \cos 2 \theta_{13} \right)
    \right]\,, \\
E_3 &\simeq \frac{\Delta m_{31}^2}{2E} \left[
    \frac{1}{2}\left(1+A+C_{13}\right) + \frac{1}{2C_{13}} \alpha \, s_{12}^2
    \left( C_{13} - 1 + A \cos 2 \theta_{13} \right)
    \right]\,, \label{eq:alpha_EV_3} 
\end{align}
where
\begin{equation}
C_{13}  \equiv  \sqrt{\sin^2 2\theta_{13} + (A-\cos 2\theta_{13})^2} \,. 
\end{equation}
To calculate the probabilities to first order in $\alpha$ we followed
two different approaches: First, we used the Cayley--Hamilton
formalism as described in \App~\ref{app:cayley-hamilton}, to series
expand the evolution matrix, and second, we considered the
constant-density limit of the expansion of the evolution equation
described in \App~\ref{app:arbitrary_alpha}. Both methods gave the
same results, which is a useful cross-check of our
calculations. Writing
\begin{equation}
P_{\alpha\beta} = P_{\alpha\beta}^{(0)} + 
\alpha \, P_{\alpha\beta}^{(1)} + {\mathcal O}( \alpha^2 ) \,, 
\end{equation}
we obtain the following expressions for the $\nu_e \to \nu_e$ channel: 
\begin{align}
P_{ee}^{(0)} &= 1 - \frac{\sin^2 2\theta_{13}}{C_{13}^2}  
   \sin^2 C_{13} \Delta \,, \label{eq:Pee0_alpha}\\
P_{ee}^{(1)} &= 2 s^2_{12} \, \frac{\sin^2 2\theta_{13}}{C_{13}^2} 
   \sin C_{13} \Delta \nonumber\\
&\times \left[ \Delta \, \frac{\cos C_{13}\Delta}{C_{13}}  (1 - A
  \cos 2\theta_{13}) - A \, \frac{\sin C_{13}\Delta}{C_{13}} \,
  \frac{\cos 2\theta_{13} - A}{C_{13}}
  \right] \,.
\end{align}
Similarly, for the $\nu_e \to \nu_\mu$ channel we find
\begin{align}
P_{e\mu}^{(0)} &= s^2_{23} \, \frac{\sin^2 2\theta_{13}}{C_{13}^2} \, 
  \sin^2 C_{13} \Delta \,, \label{eq:Pem0_alpha}\\
P_{e\mu}^{(1)} &= - 2 s^2_{12} \, s^2_{23} \, \frac{\sin^2
  2\theta_{13}}{C_{13}^2} \, \sin C_{13} \Delta \nonumber\\
&\times 
  \left[ \Delta \, \frac{\cos C_{13} \Delta}{C_{13}} (1-A \cos 2\theta_{13}) 
- A \, \frac{\sin C_{13} \Delta}{C_{13}} \, 
  \frac{\cos 2\theta_{13} - A}{C_{13}} \right] \nonumber\\
&+ s_{13} \, \sin 2\theta_{12} \sin 2\theta_{23}
  \frac{\sin C_{13} \Delta}{AC_{13}^2} \Big\{ \sin \delta_{\rm CP} \left[ \cos
  C_{13} \Delta - \cos (1+A)\Delta \right] C_{13} \nonumber\\
&+ \cos \delta_{\rm CP} \left[ C_{13} \sin (1+A)\Delta - (1-A \cos
  2\theta_{13}) \sin C_{13} \Delta \right] \Big\} \,,\label{eq:Pem1_alpha}
\end{align}
and finally, for the $\nu_\mu \to \nu_\tau$ channel we have
\begin{align}
P_{\mu\tau}^{(0)} &= \frac{1}{2} \sin^2 2\theta_{23} \bigg[
  \left( 1 - \frac{ \cos 2\theta_{13} - A}{C_{13}} \right) \sin^2
  \frac{1}{2} (1+A-C_{13})\Delta \nonumber\\
&+ \left( 1 + \frac{ \cos 2\theta_{13} -A }{C_{13}}\right)
  \sin^2 \frac{1}{2}(1+A+C_{13})\Delta - \frac{1}{2} \frac{\sin^2
  2\theta_{13}}{C_{13}^2} \sin^2 C_{13} \Delta \bigg] \,,
\label{eq:Pmt0_alpha} \\
P_{\mu\tau}^{(1)} &= - \frac{1}{2} \sin^22\theta_{23} \, \Delta
   \bigg\{ 2 \left[ c_{12}^2 - s_{12}^2 s_{13}^2 \frac{1}{C_{13}^2} (1
   + 2 s_{13}^2 A + A^2) \right] \cos C_{13} \Delta \sin  (1+A)\Delta
   \nonumber\\
&+ 2 \left[ c_{12}^2 c_{13}^2 - c_{12}^2
   s_{13}^2 + s_{12}^2 s_{13}^2 + \left( s_{12}^2 s_{13}^2 - c_{12}^2
   \right) A \right] \frac{\sin C_{13} \Delta}{C_{13}} 
   \cos (1+A) \Delta \nonumber\\
&+ s_{12}^2 \, \frac{\sin^22\theta_{13}}{C_{13}^2} \,
   \frac{\sin C_{13} \Delta}{C_{13}} \nonumber\\
&\times \left[ \frac{A}{\Delta} \, \sin (1+A)\Delta +
     \frac{A}{\Delta} \, \frac{\cos 2 \theta_{13} - A}{C_{13}} \, 
     \sin C_{13} \Delta 
     - (1 - A \cos 2 \theta_{13}) \cos C_{13} \Delta \right] \bigg\}
\nonumber\\
&+ \frac{s_{13} \, \sin 2 \theta_{12} \, \sin 2 \theta_{23}}
     {2 A c_{13}^2} \bigg\{
   2 c_{13}^2 \sin \delta_{\rm CP} \frac{\sin C_{13}
   \Delta}{C_{13}} \left[ \cos C_{13} \Delta - \cos (1+A)\Delta \right]
   \nonumber\\
&- \cos 2 \theta_{23} \cos \delta_{\rm CP} 
   (1+A) \left[ \cos C_{13} \Delta - \cos (1+A) \Delta \right]^2
   \nonumber\\
&+ \cos 2 \theta_{23} \cos \delta_{\rm CP} 
   \left[ \sin (1+A)\Delta + \frac{\cos 2 \theta_{13} - A}{C_{13}} \,
   \sin C_{13} \Delta \right] \nonumber\\
&\times \left[ 
   (1 + 2 s_{13}^2 A + A^2) \frac{\sin C_{13} \Delta}{C_{13}} - 
   (1+A) \sin (1+A)\Delta \right] \bigg\}\,.
\label{eq:Pmt1_alpha}
\end{align}
In the above formulas, one may identify the effective mixing angle in matter 
in the 1-3 sector $\theta'_{13}$, which is determined through the expressions
\begin{equation}
\sin 2\theta'_{13} = \frac{\sin 2 \theta_{13}}{C_{13}} 
\,, \qquad
\cos 2\theta'_{13} = \frac{\cos 2 \theta_{13} - A}{C_{13}} \,, 
\end{equation}
appearing frequently in
\eqs~(\ref{eq:Pee0_alpha})--(\ref{eq:Pmt1_alpha}). Furthermore, the
combination $C_{13}\Delta$ appearing as the argument of sine or cosine
corresponds to the effective $\Delta$ in matter. In the limit when
$\theta_{13}$ is small, one has $C_{13} \simeq A-1$, and expanding
\eqs~(\ref{eq:Pee0_alpha})--(\ref{eq:Pmt1_alpha}) up to second order
in $s_{13}$ yields the double expansions given in
\Sec~\ref{sec:double}, except for the terms of order $\alpha^2$.

Equations~(\ref{eq:alpha_EV_1})--(\ref{eq:alpha_EV_3}) and
(\ref{eq:Pee0_alpha})--(\ref{eq:Pem1_alpha}) have previously been
derived in \Ref~\cite{Cervera:2000kp} and confirmed in
\Ref~\cite{Freund:2001pn}. Expressions (\ref{eq:Pmt0_alpha}) and
(\ref{eq:Pmt1_alpha}) are new.

\subsection{Vacuum oscillation probabilities up to first order in 
$\mbox{\boldmath$\alpha$}$}
\label{sec:alpha_vac}

Taking the limit $A \to 0$ in 
\eqs~(\ref{eq:Pee0_alpha})--(\ref{eq:Pmt1_alpha}), it is straightforward 
to obtain the neutrino oscillation probabilities in vacuum to first 
order in $\alpha$:
\begin{align}
P_{ee}^{(0)\mathrm{vac}} &= 1 - \sin^2 2\theta_{13} \sin^2 \Delta \,,\\
P_{ee}^{(1)\mathrm{vac}} &= \Delta  
s^2_{12} \sin^2 2 \theta_{13} \sin 2 \Delta \,,\\
P_{e\mu}^{(0)\mathrm{vac}} &= \sin^2 2\theta_{13} s^2_{23} \sin^2
\Delta \,,\\
P_{e\mu}^{(1)\mathrm{vac}} &= - \Delta s^2_{12} \sin^2 2 \theta_{13}
s^2_{23} \sin 2 \Delta \nonumber \\
&+ \Delta \sin 2 \theta_{12} s_{13} c^2_{13} 
\sin 2 \theta_{23} ( 2 \sin \delta_{\mathrm{CP}} \sin^2
\Delta + \cos \delta_{\mathrm{CP}} \sin 2 \Delta ) \,,\\
P_{\mu\tau}^{(0)\mathrm{vac}} &= c^4_{13} \sin^2 2 \theta_{23}
\sin^2 \Delta \,,\\
P_{\mu\tau}^{(1)\mathrm{vac}} &= - \Delta c^2_{13} 
\sin^2 2 \theta_{23} ( c^2_{12} - s^2_{13} s^2_{12}) 
\sin 2 \Delta \nonumber \\ 
&+  \Delta \sin 2 \theta_{12} \, s_{13} c^2_{13} \, \sin 2\theta_{23} 
(2 \sin \delta_{\mathrm{CP}} \sin^2
\Delta  - \cos \delta_{\mathrm{CP}} \cos 2 \theta_{23} \sin 2 \Delta
) \,.
\end{align}

\section{Series expansion up to first order in $\mbox{\boldmath$s_{13}$}$}
\label{sec:s13}

In this section, we expand the probabilities up to first order in the
small parameter $s_{13}$ while keeping their exact dependence on
$\alpha$. These formulas are expected to be useful whenever neutrino
oscillations driven by the solar mass squared difference $\Delta
m^2_{21}$ are important. In practice, this means low neutrino energies
and long baselines.

\subsection{Matter of constant density }
\label{sec:s13_const}

The eigenvalues of the Hamiltonian (\ref{eq:ham1}) up to first order in 
$s_{13}$ are given by 
\begin{align}
E_1 &\simeq \frac{\Delta m_{31}^2}{2E} \left[ \frac{A}{2} +
    \frac{\alpha}{2}(1- C_{12}) \right] \,, \label{eq:theta_EV_1}\\
E_2 &\simeq \frac{\Delta m_{31}^2}{2E} \left[ \frac{A}{2} +
    \frac{\alpha}{2}(1+ C_{12}) \right] \,, \\
E_3 &\simeq \frac{\Delta m_{31}^2}{2E} \,, \label{eq:theta_EV_3}
\end{align}
where
\begin{equation}
C_{12} \equiv \sqrt{\sin^2 2\theta_{12} + \left( \cos
  2\theta_{12} - \frac{A}{\alpha} \right)^2} \,.
\end{equation}
Note that these eigenvalues are independent of the expansion parameter 
$s_{13}$, which is consistent with \eqs~(\ref{eq:EV1})--(\ref{eq:EV3}), where 
the lowest-order in $s_{13}$ corrections to the eigenvalues appear only at 
order $s_{13}^2$. 

As in the case of the single expansion in $\alpha$, we have calculated
the probabilities using two different methods: by series expanding the
evolution matrix using the Cayley--Hamilton formalism as described in
\App~\ref{app:cayley-hamilton}, and using the constant-density limit
of the single expansion in $s_{13}$ of the evolution equation
described in \App~\ref{app:arbitrary_s13}. We have found identical
results. Writing the probabilities as
\begin{equation}
P_{\alpha\beta} = P_{\alpha\beta}^{(0)} + s_{13} \, P_{\alpha\beta}^{(1)} 
+ {\mathcal O}(s_{13}^2) \,,
\end{equation}
we obtain the following expressions for the $\nu_e \to \nu_e$ channel: 
\begin{align}
P_{ee}^{(0)} &= 1 - \frac{\sin^2 2\theta_{12}}{C_{12}^2} \, \sin^2
  \alpha C_{12} \Delta \,, \label{eq:Pee0_theta} \\
P_{ee}^{(1)} &= 0 \,.
\end{align}
The absence of any first order corrections to $P_{ee}$ is consistent
with \eq~(\ref{eq:Pee}). Similarly, for the $\nu_e \to \nu_\mu$
channel we find
\begin{align}
P_{e\mu}^{(0)} &= c_{23}^2 \,\frac{\sin^2 2\theta_{12}}{C_{12}^2} \,
 \sin^2 \alpha C_{12} \Delta \,, \label{eq:Pem0_theta}\\
P_{e\mu}^{(1)} &= \frac{\sin 2\theta_{12}}{C_{12}} \, \sin
  2\theta_{23} \, \frac{(1-\alpha) \sin \alpha C_{12} \Delta}
  {1-A-\alpha + A\alpha c_{12}^2} 
 \bigg\{ \sin \delta_{\rm CP} \left[ \cos
  \alpha C_{12} \Delta - \cos (A+\alpha-2) \Delta \right] \nonumber\\
&- \cos \delta_{\rm CP} \left[ \sin (A+\alpha-2)\Delta 
 - \sin \alpha C_{12} \Delta \left( \frac{\cos
  2\theta_{12} - \frac{A}{\alpha}}{C_{12}} - 
  \frac{\alpha A C_{12}}{2(1-\alpha)}
  \frac{\sin^2 2\theta_{12}}{C_{12}^2} \right) \right] \bigg\} \,,
\label{eq:Pem1_theta}
\end{align}
and for the $\nu_\mu \to \nu_\tau$ channel we have
\begin{align}
P_{\mu\tau}^{(0)} &= \frac{1}{2} \sin^2 2\theta_{23} 
  \bigg[ 1 - 
  \frac{1}{2}\frac{\sin^2 2\theta_{12}}{C_{12}^2} \sin^2 \alpha C_{12} \Delta
 - \cos (\alpha C_{12} + A + \alpha - 2) \Delta 
\nonumber\\
&  - \left( 1 - \frac{\cos 2\theta_{12} - \frac{A}{\alpha}}{C_{12}}
  \right) \sin \alpha C_{12} \Delta \sin
  (A+\alpha-2) \Delta \bigg] \,, \\
P_{\mu\tau}^{(1)} &= \frac{\sin 2\theta_{12}}{C_{12}} \, \sin 2\theta_{23} \,
\frac{1}{1 - A - \alpha + A \alpha c_{12}^2} \nonumber\\
&\times \bigg\{ \frac{\alpha A C_{12}}{2} \cos 2\theta_{23} \cos
\delta_{\rm CP}
\bigg[ \left( \cos \alpha C_{12} \Delta - \cos (A + \alpha - 2) \Delta
  \right)^2 \nonumber\\
&+ \left( \frac{\cos 2\theta_{12} - \frac{A}{\alpha}}{C_{12}} \sin \alpha
  C_{12} \Delta + \sin (A + \alpha - 2) \Delta \right) \nonumber\\
&\times \left( \left( \frac{\cos 2\theta_{12} - \frac{A}{\alpha}}{C_{12}} +
  \frac{2(1-\alpha)}{\alpha AC_{12}} \right) \sin \alpha C_{12}
  \Delta + \sin (A + \alpha - 2) \Delta \right) \bigg] \nonumber\\
&+ \sin \delta_{\rm CP} \, (1-\alpha) \left( \cos \alpha C_{12}
\Delta - \cos (A + \alpha - 2) \Delta \right) \sin \alpha C_{12} \Delta \bigg\}
\,. \label{eq:Pmt1_theta}
\end{align}
In this case, one may identify in
\eqs~(\ref{eq:Pee0_theta})--(\ref{eq:Pmt1_theta}) the effective ``solar'' 
mixing angle in matter $\theta'_{12}$, which is determined by
\begin{equation}
\sin 2\theta'_{12} = \frac{\sin 2 \theta_{12}}{C_{12}} \,, \qquad
\cos 2\theta'_{12} = \frac{\cos 2 \theta_{12} - \frac{A}{\alpha}}{C_{12}} \,. 
\end{equation}
The combination $\alpha C_{12} \Delta$ appearing as argument of sine
or cosine corresponds to oscillations with the ``solar'' frequency in
matter. Furthermore, we note that in the limit when $\alpha$ is small,
one has $C_{12} \simeq A /\alpha - \cos 2\theta_{12}$, and
expanding \eqs~(\ref{eq:Pee0_theta})--(\ref{eq:Pmt1_theta}) up to
second order in $\alpha$ yields the double expansions given in
\Sec~\ref{sec:double}, except for the terms of order $s_{13}^2$.

Neutrino oscillations probabilities in matter of constant density
expanded to first order in $s_{13}$, but exact in $\alpha$, presented
in this subsection, have not been previously published and are
entirely new.

\subsection{Vacuum oscillation probabilities up to first order in 
$\mbox{\boldmath$s_{13}$}$}
\label{sec:s13_vac}

Taking the limit $A \to 0$ in
\eqs~(\ref{eq:Pee0_theta})--(\ref{eq:Pmt1_theta}), one recovers the
vacuum probabilities expanded to first order in $s_{13}$:
\begin{align}
P_{ee}^{(0)\mathrm{vac}} &= 
   1 - \sin^2 2 \theta_{12} \sin^2 \alpha \Delta \,,\\
P_{ee}^{(1)\mathrm{vac}} &= 0 \,,\\
P_{e\mu}^{(0)\mathrm{vac}} &=  
\sin^2 2 \theta_{12} c_{23}^2 \sin^2 \alpha \Delta \,,\\
P_{e\mu}^{(1)\mathrm{vac}} &=  \cos \delta_{\mathrm{CP}} \sin 2
   \theta_{12} \sin 2 
\theta_{23} [\sin^2 \Delta - \sin^2 (1-\alpha) \Delta +
\cos 2 \theta_{12} \sin^2 \alpha \Delta] \nonumber \\ &+\frac{1}{2}
\sin \delta_{\mathrm{CP}} \sin 2 \theta_{12} \sin 2
\theta_{23}  [-\sin 2 \Delta + \sin 2
(1-\alpha) \Delta + \sin 2 \alpha \Delta] \,,\\
P_{\mu\tau}^{(0)\mathrm{vac}} &= s_{12}^2 \sin^2 2 \theta_{23}
\sin^2 \Delta + c_{12}^2 \sin^2 2 \theta_{23}
[\sin^2(1-\alpha)\Delta - s_{12}^2 \sin^2 \alpha \Delta ] \,,\\
P_{\mu\tau}^{(1)\mathrm{vac}} &= \cos \delta_{\mathrm{CP}} \sin 2 \theta_{12}
\sin 2 \theta_{23} \cos 2 \theta_{23} [-\sin^2 \Delta + \sin^2
(1-\alpha) \Delta + \cos 2 \theta_{12} \sin^2 \alpha \Delta ]
\nonumber \\ &+ \frac{1}{2} \sin \delta_{\mathrm{CP}} \sin 2
\theta_{12} \sin 2 \theta_{23} [-\sin 2 \Delta + \sin
2(1-\alpha)\Delta + \sin 2 \alpha \Delta ]\,.
\end{align}

\section{Qualitative discussion and tests of accuracy}
\label{sec:quality}

In this section, we discuss the qualitative behavior of the neutrino
oscillation probabilities and the relevance of matter effects
(\Sec~\ref{sec:Pemu}). We also assess quantitatively the accuracy of
the various expansions and compare each type of expansion with an
exact numerical calculation of the corresponding probability
(\Secs~\ref{sec:comparing} and \ref{sec:double-acc}).  We examine in
detail which type of expansion is most accurate, depending on the
values of the fundamental parameters (especially $\alpha$ and
$s_{13}$), and on the experimental configuration characterized by the
neutrino energy $E$ and the baseline length $L$. In
\Sec~\ref{sec:experiments}, we discuss issues related to the
application of our formulas to neutrino oscillation experiments.  As
we have shown in \Sec~\ref{sec:notation}, the probabilities for all
oscillation channels can be obtained from the expressions for
$P_{e\mu}$ and $P_{\mu\tau}$ [see
\eqs~(\ref{eq:Pee_gen})--(\ref{eq:relations})].  Therefore, we focus
in the following on the $\nu_e\to\nu_\mu$ channel; we briefly comment
also on the other neutrino oscillation channels.

\subsection{The relevance of matter effects and the probability
  \boldmath$P_{e\mu}$}
\label{sec:Pemu}

\begin{figure}[t]
    \centering
    \includegraphics[width=0.7\linewidth]{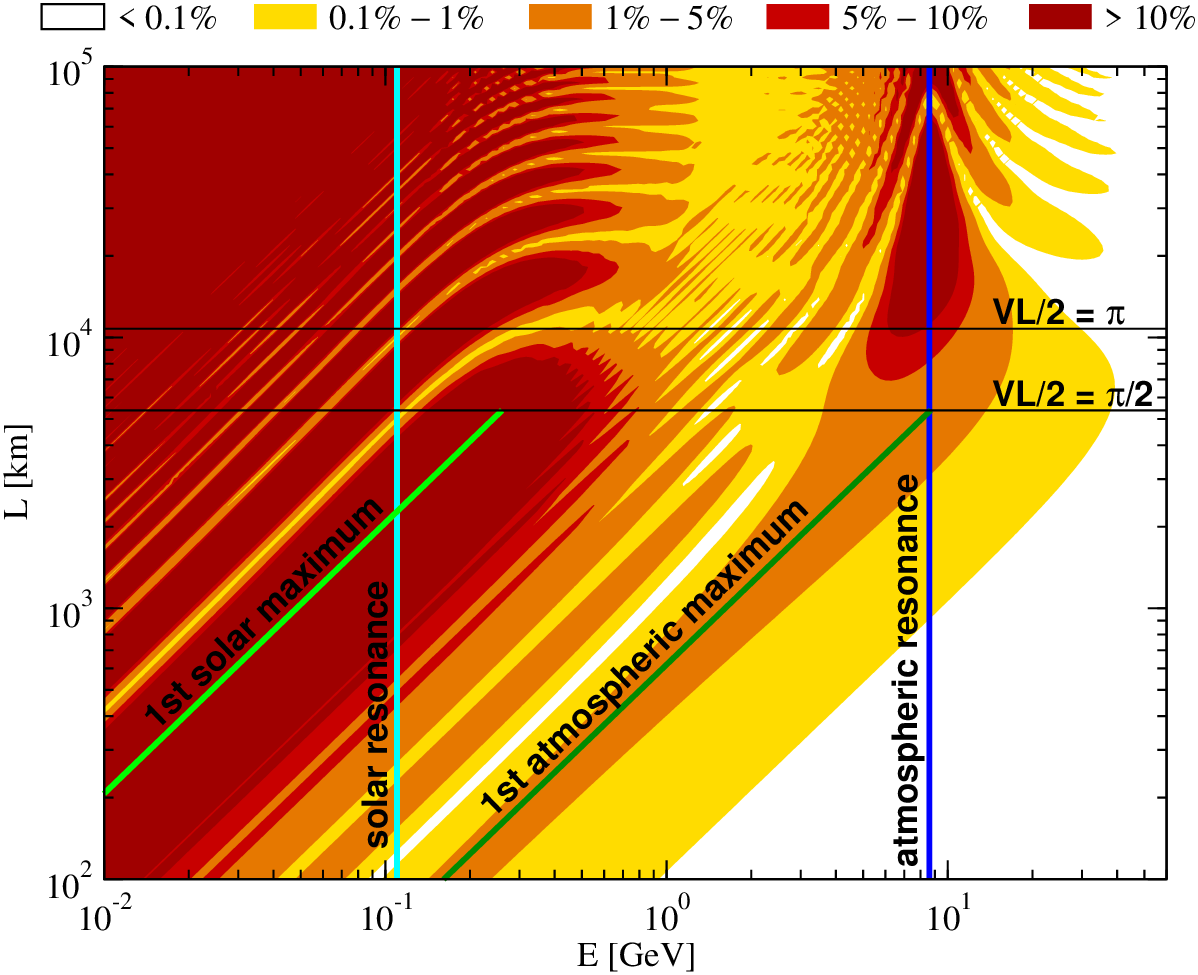}
    \caption{\label{fig:P_contours} Contours of $P_{e\mu}$ calculated
    numerically without approximations for $\sin^2 2\theta_{13} =
    0.02$, $\Delta m_{31}^2 = 2 \cdot 10^{-3} \, {\rm eV}^2$, $\alpha
    = 0.03$, $\theta_{12} = 33^\circ$, $\theta_{23}= 45^\circ$, and
    $\delta_{\rm CP} = 0$. Matter of constant density $\rho = 3 \,
    {\rm g/cm^3}$ is assumed, and the probability is averaged over a
    Gaussian energy resolution of 1\%.}
\end{figure}

Before considering the accuracy of the expansion formulas, we discuss 
in this subsection the features and the relevance of matter effects
for the $\nu_e\to\nu_\mu$ oscillation probability $P_{e\mu}$. The
discussion applies also to $P_{e\tau}$; matter effects on the probabilities 
$P_{\mu\mu}$, $P_{\tau\tau}$, and $P_{\mu\tau}$ are significantly weaker  
than those on $P_{e\mu}$ and $P_{e\tau}$. 
Figure~\ref{fig:P_contours} shows contours of $P_{e\mu}$ for 
matter of constant density calculated numerically from \eq~(\ref{eq:ham1}) 
without any approximations, for a wide range of baseline lengths and 
neutrino energies.
Many features of this figure can be understood by considering the expression 
for the two-flavor neutrino oscillation probability in matter of 
constant density:
\begin{equation}
\label{eq:P_twonu}
P = \frac{\sin^2 2\theta}{C^2} \sin^2  C \Delta_{2\nu} \,,\qquad
C \equiv \sqrt{\sin^2 2\theta + (\cos 2 \theta - A_{2\nu})^2} \,,
\end{equation}
where, in analogy to \eqs~(\ref{eq:def_Delta}) and (\ref{eq:def_A}), we
define $\Delta_{2\nu} \equiv \Delta m^2 L /(4 E)$, $A_{2\nu} \equiv 2
E V / \Delta m^2$, and $\theta$ and $\Delta m^2$ are the generic
two-flavor neutrino oscillation parameters. {}From
\eq~(\ref{eq:P_twonu}) it is clear that for $A_{2\nu} \ll 1$ one has
$C \simeq 1$, and vacuum neutrino oscillations with $P\simeq \sin^22
\theta \sin^2\Delta_{2\nu}$ are recovered. For $A_{2\nu} = \cos 2
\theta$ one can see in \eq~(\ref{eq:P_twonu}) also the MSW
resonance~\cite{MSW}, which leads to $C = \sin 2 \theta$ in 
\eq~(\ref{eq:P_twonu}) and to the effective mixing angle in matter 
$\sin^22\theta' = \sin^22\theta / C^2 = 1$.  The general resonance 
conditions for three flavors are much more complicated than those in the 
two-flavor case. However, due to the hierarchy of the mass squared 
differences and smallness of $\theta_{13}$, one can use, to a very good 
approximation, the respective effective two-flavor picture. This leads 
to the ``solar'' and ``atmospheric'' resonance energies shown as vertical 
lines in \fig~\ref{fig:P_contours}
\begin{equation}
\frac{A}{\alpha} = \frac{2 E V}{\Delta m^2_{21} } = \cos 2\theta_{12} \,,
\qquad
A = \frac{2 E V}{\Delta m^2_{31} } = \cos 2\theta_{13} \,.
\end{equation}
The ``atmospheric'' resonance is clearly visible in
\fig~\ref{fig:P_contours} at $E_\mathrm{res} \simeq 8.6$~GeV and $L
\gtrsim 5500$~km. The ``solar'' resonance occurs at the energy
$E_\mathrm{res}\simeq 0.11$~GeV. Note that the solar resonance is not
as pronounced in \fig~\ref{fig:P_contours} as the atmospheric one,
since the neutrino oscillation amplitude in vacuum
$\sin^22\theta_{12}$ is already quite large.

Far enough to the left of the vertical lines in
\fig~\ref{fig:P_contours} marking the solar and atmospheric resonance
energies one has vacuum neutrino oscillations with the ``solar''
parameters $\theta_{12}$, $\Delta m^2_{21}$ and with the
``atmospheric'' parameters $\theta_{13}$, $\Delta m^2_{31}$,
respectively. Indeed, the typical $L/E$ pattern is clearly visible in
\fig~\ref{fig:P_contours} to the left of the vertical lines. 
The
diagonal lines indicate the constant values of $L/E$ corresponding to
the first solar and atmospheric oscillation maxima in vacuum, given by
the conditions
\begin{equation}
\alpha\Delta = \frac{\Delta m_{21}^2 L}{4E} = \frac{\pi}{2} \,,
\qquad
\Delta = \frac{\Delta m_{31}^2 L}{4E} = \frac{\pi}{2} \,,
\end{equation}
respectively. To the right of the vertical lines the probability is
dominated by matter effects. For $A_{2\nu} \gg 1$ we have $C \simeq
A_{2\nu}$ in \eq~(\ref{eq:P_twonu}), which has two consequences:
First, since $A_{2\nu}\Delta_{2\nu} = VL/2$, the oscillation frequency
becomes independent of energy, and second, the oscillation amplitude
becomes suppressed at high energies because $\sin^2 2\theta/A_{2\nu}^2
\propto 1/E^2$. Both these effects are apparent in
\fig~\ref{fig:P_contours} for neutrino oscillations with the solar as
well as the atmospheric frequency.

In addition to the dependence on the neutrino energy, matter effects
depend crucially on the baseline. From \eq~(\ref{eq:P_twonu}) one 
can see that the two-flavor oscillation probabilities approach the 
vacuum ones for $C\Delta_{2\nu}\ll \pi/2$.  
Far above the MSW resonance energy, \ie\, for $A_{2\nu}\gg 1$, one
finds $C\simeq A_{2\nu}$. Hence, the short-baseline limit is
$C\Delta_{2\nu} \simeq A_{2\nu}\Delta_{2\nu} = VL/2 \ll \pi/2$,
leading to vanishing matter effects for $L \ll L_\mathrm{mat} \equiv
\pi/V \simeq 5453$~km.  The horizontal lines in
\fig~\ref{fig:P_contours} indicate this baselines of the first
``matter effect maximum'' for large energies at $L = L_\mathrm{mat}$,
and the first ``matter effect minimum'', $L= 2L_\mathrm{mat}\simeq
10907$~km. At the baseline $2L_\mathrm{mat}$, where $A\Delta = VL/2 =
\pi$, the terms proportional to $\alpha^2$ and $\alpha s_{13}$ in the
double expansion \eq~(\ref{eq:Pem}) disappear, and only the term
proportional to $s_{13}^2$ survives. Therefore, this baseline is
especially useful to measure $s_{13}$, since ambiguities due to
parameter degeneracies are avoided \cite{Barger:2001yr}. A dedicated
analysis of this ``magic baseline'' can be found in
\Ref\cite{Huber:2003ak}.\footnote{It should be noted that the quoted
number for $L_\mathrm{mat}$ holds for a constant matter density of
$\rho = 3 \, {\rm g/cm^3}$. It will differ for larger matter densities
like those in the mantle of the Earth, or even more drastically in
astrophysical or cosmological applications of neutrino oscillations.
In addition, the approximation of constant matter density is then
frequently not justified. For a realistic Earth matter density profile
one finds $2L_\mathrm{mat} \simeq 7250$~km~\cite{Huber:2003ak}.}  Note
that the distance $L_\mathrm{mat}$ depends only on the matter
potential $V$, \ie\ the energy $E$ as well as neutrino masses and
mixings do not enter.

Let us now consider the energy region well below the MSW resonance
energy, where \mbox{$A_{2\nu}\ll 1$}. The short-baseline behavior of
the oscillation probability in this region can be understood in the
two-flavor picture by expanding \eq~(\ref{eq:P_twonu}) for fixed
$\Delta_{2\nu}$ (\ie, fixed $L/E$) assuming $A_{2\nu} \ll 1$ and
$\Delta_{2\nu}A_{2\nu} \ll 1$.  One finds
\begin{equation} \label{eq:fingers}
P^\mathrm{matter} - P^\mathrm{vacuum} \simeq
2 \sin^2 2\theta \cos 2\theta \: \sin \Delta_{2\nu}
  \left(\sin\Delta_{2\nu} - \Delta_{2\nu} \cos\Delta_{2\nu}\right)
  A_{2\nu}\,.
\end{equation}
This relation shows that $P^\mathrm{matter}-P^\mathrm{vacuum}$
decreases linearly with $A_{2\nu} \propto E$, which means that the matter
effects vanish for very small energies, as mentioned above. In
addition, for a fixed energy the right-hand side of
\eq~(\ref{eq:fingers}) becomes zero for $\Delta_{2\nu}\ll 1$ or,
equivalently, $L \ll L_\mathrm{vac} \equiv 4\pi E/\Delta m^2$. 
Therefore, the two conditions for the smallness of matter effects due
to the short baseline can be written for all energies as $L\ll
L_\mathrm{min} \equiv \mathrm{min}(L_\mathrm{vac}, L_\mathrm{mat})$.

\begin{figure}[t]
    \centering
    \includegraphics[angle=0,width=0.9\linewidth]{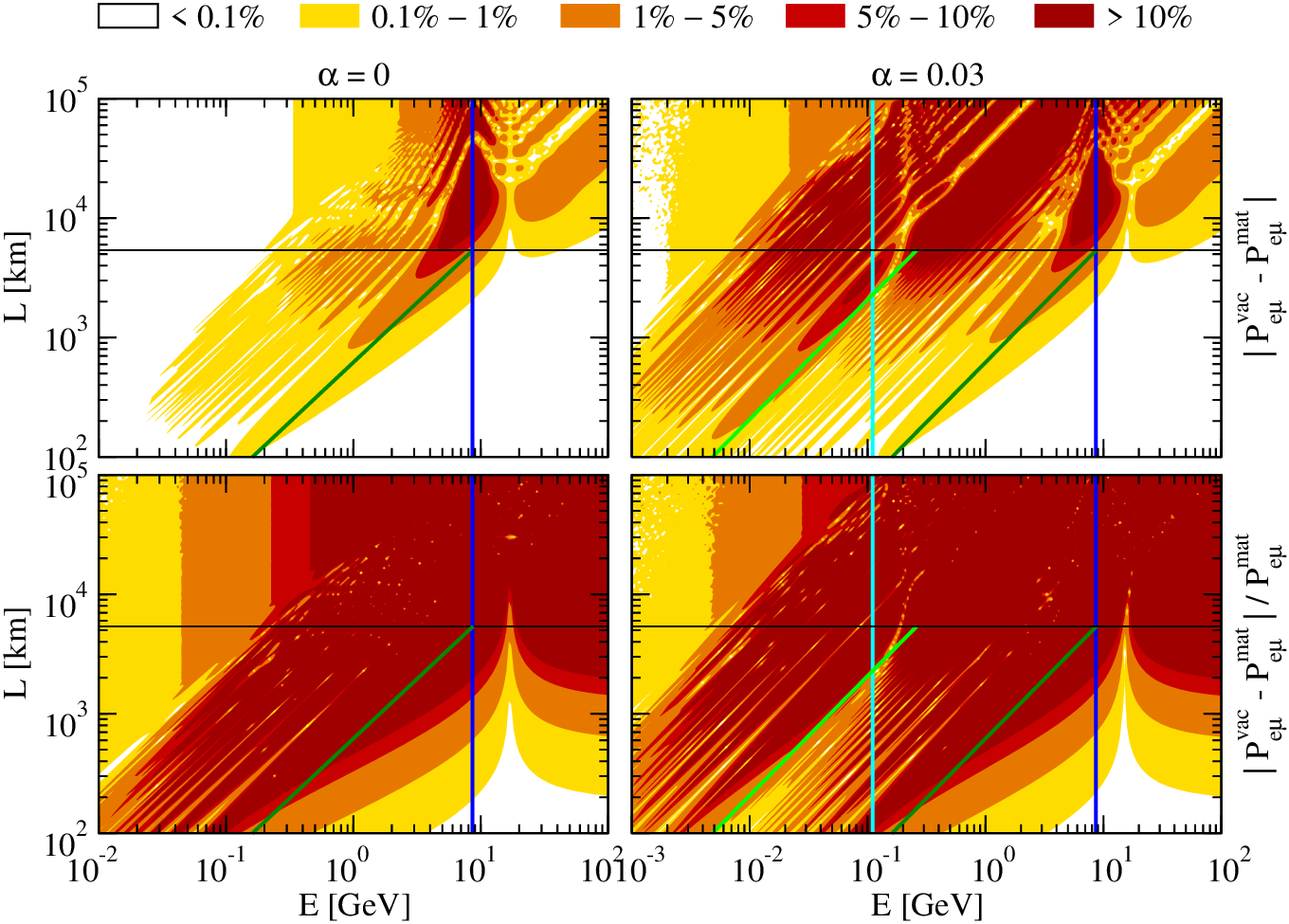}
    \caption{\label{fig:P_diff} Contours of $|P_{e\mu}^\mathrm{matter}
    - P_{e\mu}^\mathrm{vacuum}|$ (upper panels), and
    $|P_{e\mu}^\mathrm{matter} - P_{e\mu}^\mathrm{vacuum}| /
    P_{e\mu}^\mathrm{matter}$ (lower panels), where
    $P_{e\mu}^\mathrm{matter}$ ($P_{e\mu}^\mathrm{vacuum}$) is the exact
    oscillation probability in matter (vacuum) for
    $\sin^22\theta_{13}=0.05$, $\Delta m_{31}^2 = 2 \cdot 10^{-3} \,
    {\rm eV}^2$, and $\theta_{23}= 45^\circ$. In the left panels
    $\alpha = 0$, whereas in the right panels $\alpha = 0.03$,
    $\theta_{12} = 33^\circ$, and $\delta_{\rm CP} = 0$. 
    Matter of constant density $\rho = 3 \,
    {\rm g/cm^3}$ is assumed, and the probabilities are averaged over a
    Gaussian energy resolution of 2\%. The straight lines indicate the
    relevant resonance energies, first oscillation maxima, and the first
    matter effect maximum (see also \Fig~\ref{fig:P_contours}).}
\end{figure}

The relevance of matter effects is illustrated in \fig~\ref{fig:P_diff}.  
In the upper left panel we show the probability difference 
$|P_{e\mu}^\mathrm{matter} - P_{e\mu}^\mathrm{vacuum}|$ for the case 
$\alpha=0$, such that the oscillations due to the ``solar'' frequency are 
switched off and only matter effects are present which are related to the 
``atmospheric'' resonance energy at 8.6~GeV. It can be seen that the 
difference, as well as the relative difference (lower left panel), vanish 
for $L\ll L_\mathrm{min}$.
The left plots of \fig~\ref{fig:P_diff} show that the differences vanish also 
for small energies and long baselines. This can be understood as follows. 
For $E\ll E_{\rm res}$ the oscillation amplitude in matter is always
close to that in vacuum. For very short baselines, this is also true
for the oscillation phases. At intermediate baselines, the
matter-induced correction to the oscillation phase, though much
smaller than the main ``vacuum'' term, can become comparable to unity
and so cannot be ignored. However, at long baselines $L \gg
L_\mathrm{vac}$ the averaging regime sets in, which implies that any
shifts in the oscillation phase become unimportant and the vacuum
oscillations limit is regained.

The above discussion describes the regions where
$|P_{e\mu}^\mathrm{matter} - P_{e\mu}^\mathrm{vacuum}|$ vanishes or
becomes small for $\alpha=0$ in the left plots of
\fig~\ref{fig:P_diff}.  One can also understand in this way why the
difference $|P_{e\mu}^\mathrm{matter}-P_{e\mu}^\mathrm{vacuum}|$
disappears more slowly towards low $E$ and $L$ along the lines of 
constant $L/E$ in the upper left plot of \fig~\ref{fig:P_diff}. The 
structures seen in the plot emerge from the matter dependent phase
shifts of oscillation probabilities with more or less equal
amplitudes.  This behavior can be understood in the two-flavor
picture from \eq~(\ref{eq:fingers}), which implies that
$P^\mathrm{matter}-P^\mathrm{vacuum}$ grows linearly with $E$ along
the lines of constant $\Delta_{2\nu}$ or, equivalently, constant
$L/E$. This growth is modulated in the $\Delta_{2\nu}$-direction by
the factor $\sin \Delta_{2\nu} \left(\sin\Delta_{2\nu} - \Delta_{2\nu}
\cos\Delta_{2\nu}\right)$, which describes nicely the details of the
structures extending along lines of constant $\Delta_{2\nu}$ towards
lower energies. 

The right plots of \fig~\ref{fig:P_diff} show the probability
difference $|P_{e\mu}^\mathrm{matter}-P_{e\mu}^\mathrm{vacuum}|$ and
the relative difference for the case $\alpha=0.03$. This figure
contains the structures stemming from the atmospheric resonance
energy, which were already present in the $\alpha=0$ case (left
plots). In addition, similar structures show up around the solar
resonance energy, reaching to lower energies and, due to the larger
effective mixing angle, also to shorter baselines. The right plots of
\fig~\ref{fig:P_diff} show that in the realistic case of two mass
squared differences matter effects are rather important and have to be
taken into account in a large domain of the physically interesting
parameter space.  

\subsection{Comparing the accuracy of the three types of expansions}
\label{sec:comparing}

\begin{figure}[t]
    \centering
    \includegraphics[width=0.8\linewidth]{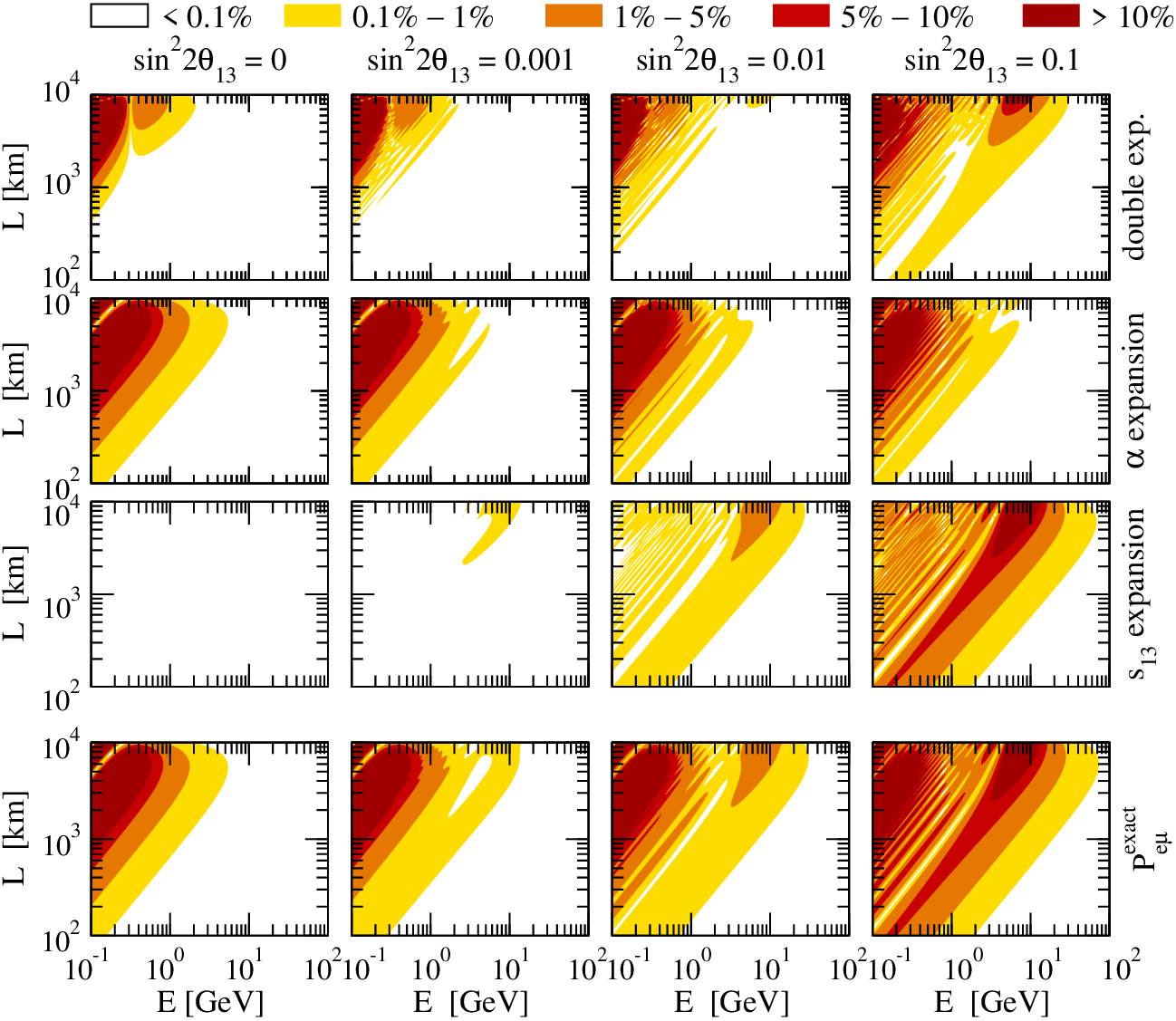}
    \caption{\label{fig:compare-abs} Rows 1, 2, and 3: absolute errors
    of the three types of expansions for $P_{e\mu}$ as functions of
    neutrino energy $E$ and baseline length $L$, for $\Delta m_{31}^2
    = 2 \cdot 10^{-3} \, {\rm eV}^2$, $\alpha = 0.03$, $\theta_{12} =
    33^\circ$, $\theta_{23}= 45^\circ$, $\delta_{\rm CP} = 0$, and
    several values of $\sin^22\theta_{13}$. Matter of constant density
    $\rho = 3 \, {\rm g/cm^3}$ is assumed. Row 4 shows contours of
    $P_{e\mu}$ calculated numerically.}
\end{figure}

\begin{figure}[t]
    \centering
    \includegraphics[width=0.8\linewidth]{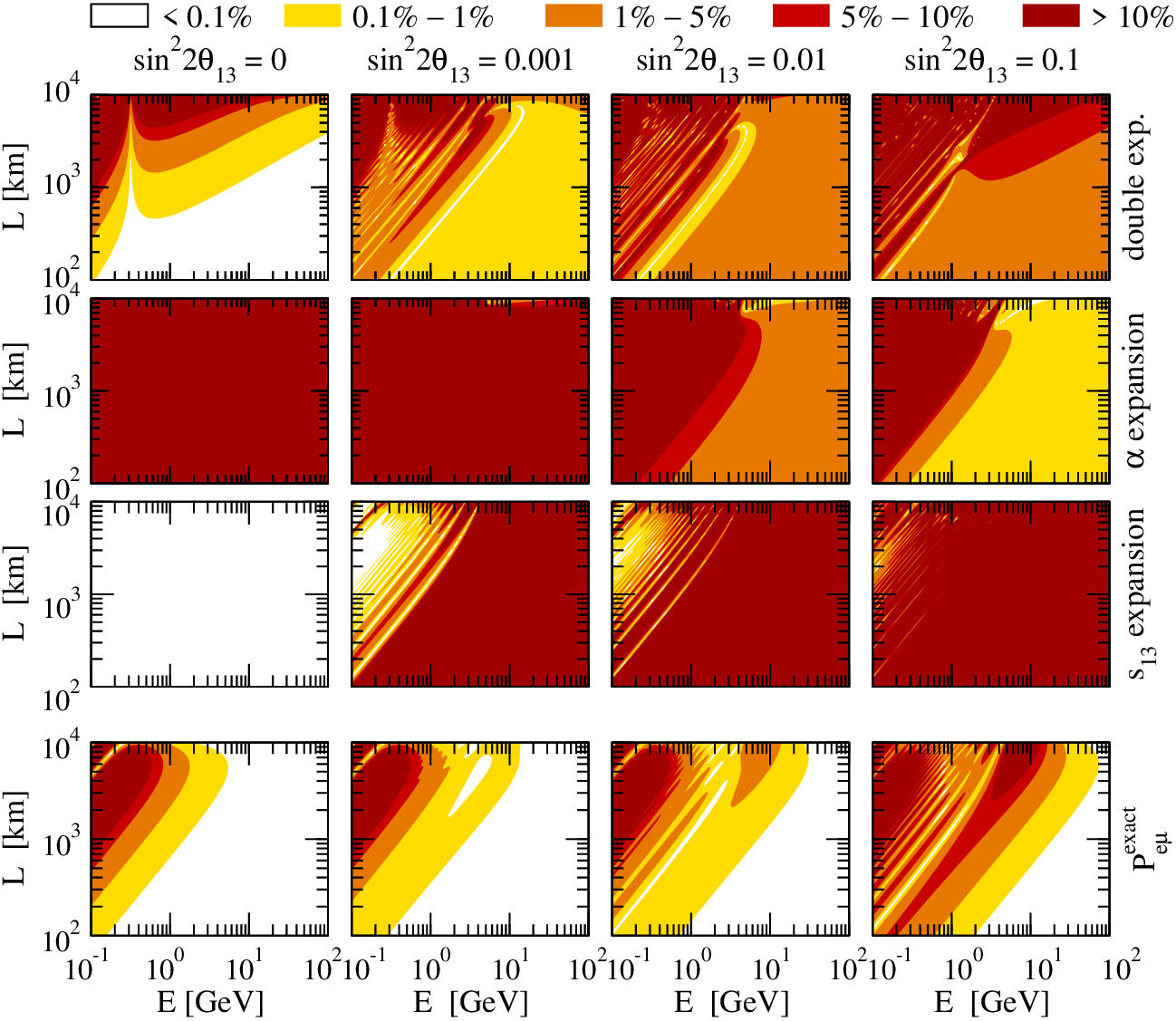}
    \caption{\label{fig:compare-rel} Rows 1, 2, and 3: relative errors
    of the three types of expansions for $P_{e\mu}$ as functions of
    neutrino energy $E$ and baseline length $L$, for $\Delta m_{31}^2
    = 2 \cdot 10^{-3} \, {\rm eV}^2$, $\alpha = 0.03$, $\theta_{12} =
    33^\circ$, $\theta_{23}= 45^\circ$, $\delta_{\rm CP} = 0$, and
    several values of $\sin^22\theta_{13}$. Matter of constant density
    $\rho = 3 \, {\rm g/cm^3}$ is assumed. Row 4 shows contours of
    $P_{e\mu}$ calculated numerically.}
\end{figure}

In order to test the accuracy of our analytic expressions, we shall
now compare the values of $P_{e\mu}$ obtained from the expansion
formulas \eq~(\ref{eq:Pem}) for the double expansion,
\eqs~(\ref{eq:Pem0_alpha}) and (\ref{eq:Pem1_alpha}) for the single
expansion in $\alpha$, and \eqs~(\ref{eq:Pem0_theta}) and
(\ref{eq:Pem1_theta}) for the single expansion in $s_{13}$ with the
exact values of $P_{e\mu}$ calculated numerically.
Figure~\ref{fig:compare-abs} shows the absolute errors
$|P_{e\mu}^\mathrm{expansion} - P_{e\mu}^\mathrm{exact}|$, and
\fig~\ref{fig:compare-rel}, the relative errors
$|P_{e\mu}^\mathrm{expansion} -
P_{e\mu}^\mathrm{exact}|/P_{e\mu}^\mathrm{exact}$ for the three types
of expansions as functions of the neutrino energy $E$ and baseline
length $L$ for various values of $\sin^22\theta_{13}$. For reference,
we display in the lowest row of graphs in each of these figures also
the probability $P_{e\mu}$ itself, calculated numerically without any
approximations.

First, we observe that in general the absolute errors shown in
\fig~\ref{fig:compare-abs} are rather small---at the 0.1\% level in a
large part of the parameter space, whereas the relative errors shown
in \fig~\ref{fig:compare-rel} are considerably larger, due to the
smallness of the probability itself.  Next, we note that the series
expansions in $\alpha$, \ie, the double expansion in $\alpha$ and
$s_{13}$ and the single expansion in $\alpha$, are only valid for
\begin{equation}
\label{eq:atm_region} 
\alpha \Delta = \frac{\Delta m_{21}^2 L}{4E} \ll 1\,,
\qquad\mbox{or}\qquad 
\frac{L}{E} \ll  {10^4 \; \mathrm{km}/\mathrm{GeV}}  \,, 
\end{equation} 
\ie, far below the first solar maximum. The obvious reason is that
expanding terms of the type $\sin \alpha\Delta$ is only valid if
$\alpha\Delta$ is small, and hence, these two types of expansion
cannot account for neutrino oscillations with the ``solar''
frequency. Note that for $s_{13} = 0$ the single expansion in $\alpha$
gives $P_{e\mu} = 0$, since in that case the lowest-order term of
$P_{e\mu}$ is proportional to $\alpha^2$, and our single expansion in
$\alpha$ only contains terms up to first order in $\alpha$.  This
explains why at very small values of $s_{13}$ the double expansion
(which includes terms of second order in $\alpha$) is more accurate
than the single expansion in $\alpha$.
In contrast, the neutrino oscillations with the ``solar'' frequency
(for which the condition in \eq~(\ref{eq:atm_region}) is violated) are
in general very well accounted for by the single expansion in
$s_{13}$, since it retains the exact dependence of the probability on
$\alpha$.  This expansion is the best one for relatively small values
of $s_{13}$ and large values of $L/E$.

Figures~\ref{fig:compare-abs} and \ref{fig:compare-rel} allow us to
put the above observations on a more quantitative basis. In the region
of $L/E$ given in \eq~(\ref{eq:atm_region}) (lower-right parts of the
graphs), where the oscillations are mainly driven by the
``atmospheric'' mass squared difference $\Delta m^2_{31}$, the double
expansion and the single expansion in $\alpha$ work rather well. For
not too large values of $\sin^22\theta_{13}$ the double expansion is
most accurate, with absolute errors at the 0.1\% level and relative
errors not exceeding 1\% (5\%) for $\sin^22\theta_{13} = 0.001 \,
(0.01)$. However, for values of $\sin^2 2\theta_{13}$ close to the
current upper bound the single expansion in $\alpha$ becomes better,
with relative errors smaller than 1\%. The reason for this is that for
large values of $\sin^22\theta_{13}$ the neutrino oscillations driven
by the ``atmospheric'' frequency and mixing angle $\theta_{13}$
completely dominate the probability, and, in addition, for relatively
large values of $L$ and energies close to 10 GeV, the atmospheric
resonance becomes important. Since the single expansion in $\alpha$ is
exact in $s_{13}$, it describes the case of relatively large $s_{13}$
well, and the atmospheric resonance is also correctly accounted
for. At the same time, the accuracy of the double expansion becomes
slightly worse near the first atmospheric maximum and the
resonance. As was pointed out in \Ref\cite{Freund:2001pn}, for large
values of $s_{13}$ the accuracy of the double expansion can be improved by
replacing the term proportional to $s_{13}^2$ in \eq~(\ref{eq:Pem}) by
the term given in \eq~(\ref{eq:Pem0_alpha}) as the zeroth order term
in the single expansion in $\alpha$. For $\alpha = 0$ this term describes the
probability $P_{e\mu}$ exactly to all orders in $s_{13}$.

As can be seen from \figs~\ref{fig:compare-abs} and
\ref{fig:compare-rel}, the single expansion in $s_{13}$ gives a rather
poor description of the region of $L/E$ defined in
\eq~(\ref{eq:atm_region}). The reason is that with only terms of first
order in $s_{13}$ it is not possible to obtain a correct description
of neutrino oscillations driven by $\Delta m^2_{31}$ and $\theta_{13}$. 
This is also reflected by the fact that the lowest-order in $s_{13}$ terms 
in the eigenvalues of the Hamiltonian are ${\cal O} (s_{13}^2)$ [see
\eqs~(\ref{eq:EV1})--(\ref{eq:EV3})]. On the other hand, the single
expansion in $s_{13}$ is rather accurate for large values of $L/E$
violating the condition (\ref{eq:atm_region}) (upper-left parts of the
graphs) and relatively small values of $\theta_{13}$: for $\sin^2
2\theta_{13}<0.001$ the accuracy is typically better than 1\%.

\begin{figure}[t]
    \centering 
    \includegraphics[width=0.8\linewidth]{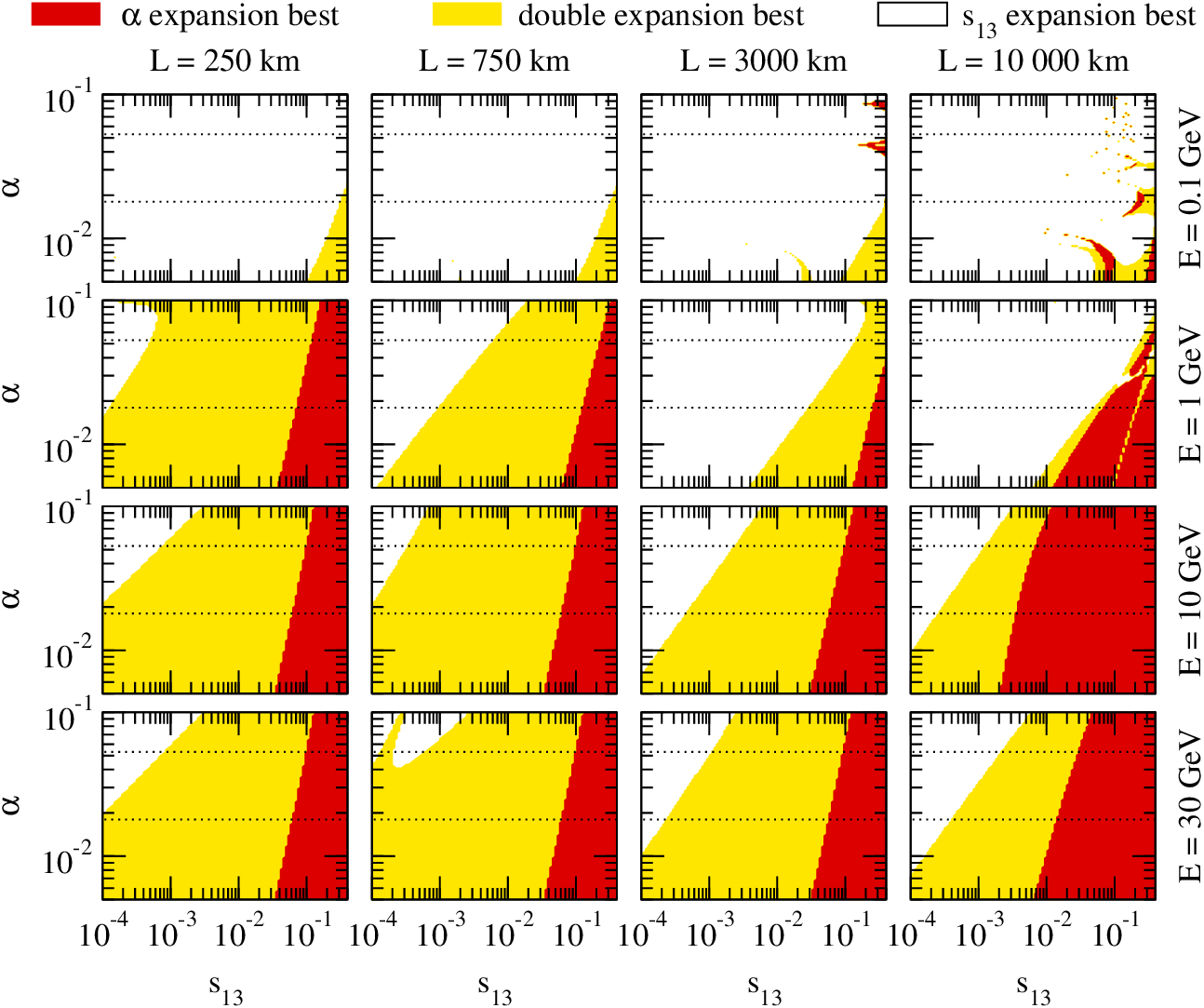}
    \caption{\label{fig:which_best}The plot shows which type of
     expansion for $P_{e\mu}$ is most accurate, depending on  
     $\alpha$ and $s_{13}$ and for different values of neutrino 
     energy $E$ and baseline length $L$. The values of the fundamental 
     neutrino parameters used are $\Delta m_{31}^2 = 2 \cdot 10^{-3} \,
     {\rm eV}^2$, $\theta_{12} = 33^\circ$, $\theta_{23}= 45^\circ$,
     and $\delta_{\rm CP} = 0$. Matter of constant density $\rho = 3 \, 
     {\rm g/cm^3}$ is assumed. The dotted lines indicate
     the $3\sigma$ allowed range of the parameter $\alpha$.}
\end{figure}
 
To conclude this subsection, we show in \fig~\ref{fig:which_best} which
type of expansion provides the most accurate expression for $P_{e\mu}$,  
depending on the values of the expansion parameters $\alpha$ and $s_{13}$ 
and for a number of fixed values of neutrino energy $E$ and  baseline 
length $L$. These plots change very little when the fundamental neutrino 
parameters $\Delta m_{31}^2$, $\theta_{12}$, $\theta_{23}$, and 
$\delta_{\rm CP}$ are varied within their allowed ranges, and also when 
one switches over to the other neutrino oscillation channels. 
As expected, one observes as a general trend that the single expansion in 
$\alpha$ is best for small $\alpha$ and large $s_{13}$, whereas the single 
expansion in $s_{13}$ is best for small $s_{13}$ and large $\alpha$. The 
double expansion is most accurate in a region where $\alpha$ and $s_{13}$ are 
of comparable order of magnitude.

In agreement with the discussion related to
\figs~\ref{fig:compare-abs} and \ref{fig:compare-rel}, we find from
\fig~\ref{fig:which_best} that in the low-energy regime $E\sim
0.1$~GeV the single expansion in $s_{13}$ is most accurate in almost
the entire $\alpha$-$\theta_{13}$ plane.  For values of $L/E$
satisfying the condition (\ref{eq:atm_region}) and energies larger
than a few GeV the double expansion is most accurate in a large
fraction of the $\alpha$-$\theta_{13}$ plane; only for values of
$s_{13}$ close to the current upper bound does the single expansion in
$\alpha$ become better. Furthermore, we note that for $E \sim 10$~GeV
and $L \gtrsim 5500$~km the atmospheric resonance is important, which
leads to a better accuracy of the single expansion in $\alpha$ (see
the rightmost panel in the third row of \fig~\ref{fig:which_best}).

We conclude that the double expansion in both $\alpha$ and $s_{13}$ is most 
accurate in a wide region of the parameter space, where $\alpha$ and 
$s_{13}$ are roughly of the same order of magnitude. The single 
expansion in $s_{13}$ has to be used whenever neutrino oscillations with the 
solar frequency are important (\eg, in the low-energy regime), whereas the 
single expansion in $\alpha$ is most accurate for values of $s_{13}$ close to 
the upper bound, or in cases where the atmospheric resonance is important.

\subsection{Accuracy of the double expansion}
\label{sec:double-acc}

\begin{figure}[t]
    \centering \includegraphics[width=0.8\linewidth]{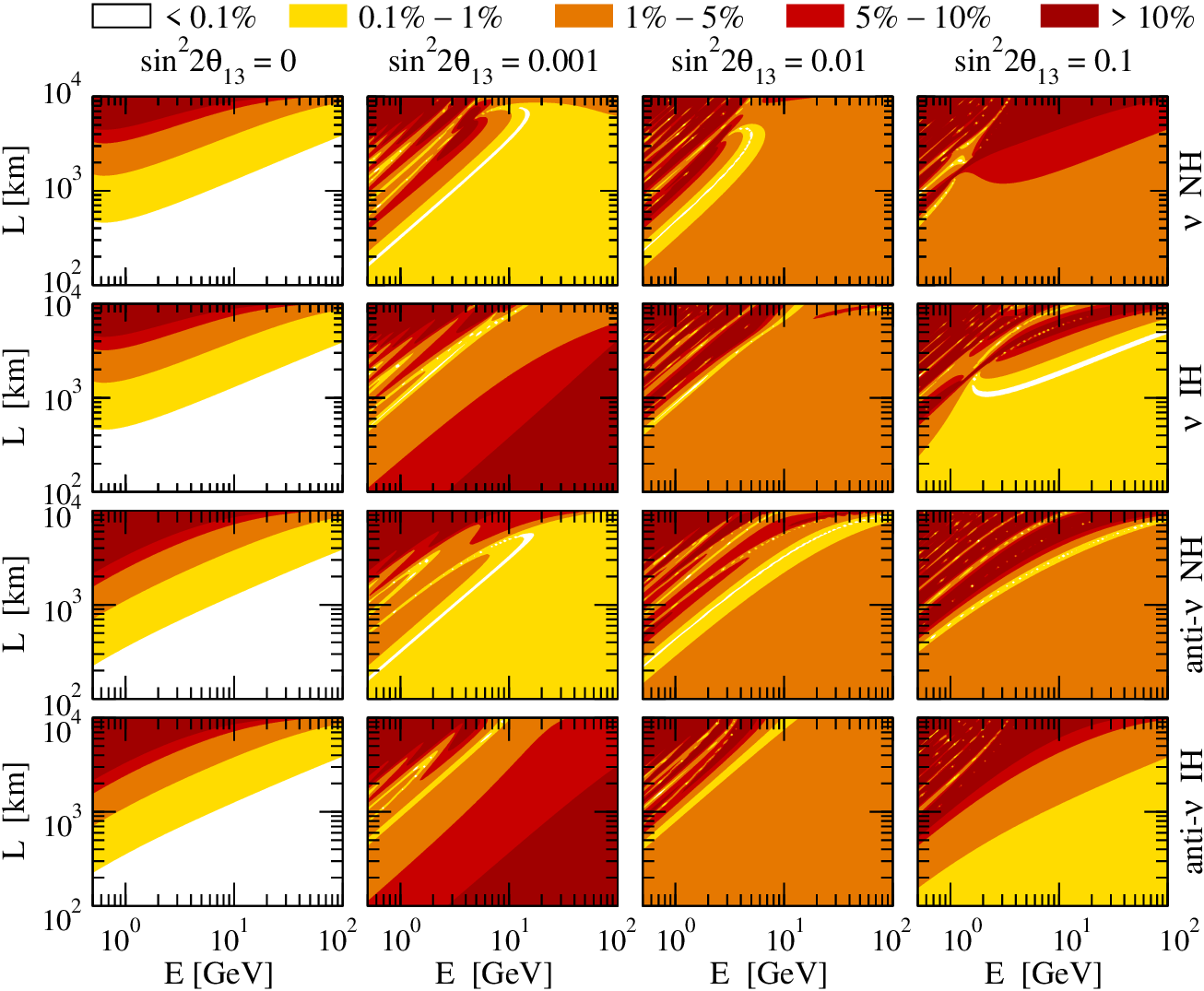}
    \caption{\label{fig:double-Pemu} Relative errors of the
    double expansion for $P_{e\mu}$ for neutrinos (rows 1 and 2), and 
    $P_{\overline{e}\overline{\mu}}$ for antineutrinos (rows 3 and 4),
    for normal hierarchy (NH) and inverted hierarchy (IH). The values of 
    the fundamental neutrino parameters used are $|\Delta m_{31}^2|
    = 2 \cdot 10^{-3} \, {\rm eV}^2$, $\alpha = 0.03$, $\theta_{12} =
    33^\circ$, $\theta_{23}= 45^\circ$, $\delta_{\rm CP} = 0$, and
    several values of $\sin^22\theta_{13}$. Matter of constant 
    density $\rho = 3 \, {\rm g/cm^3}$ is assumed.}
\end{figure}

\begin{figure}[t]
    \centering \includegraphics[width=0.8\linewidth]{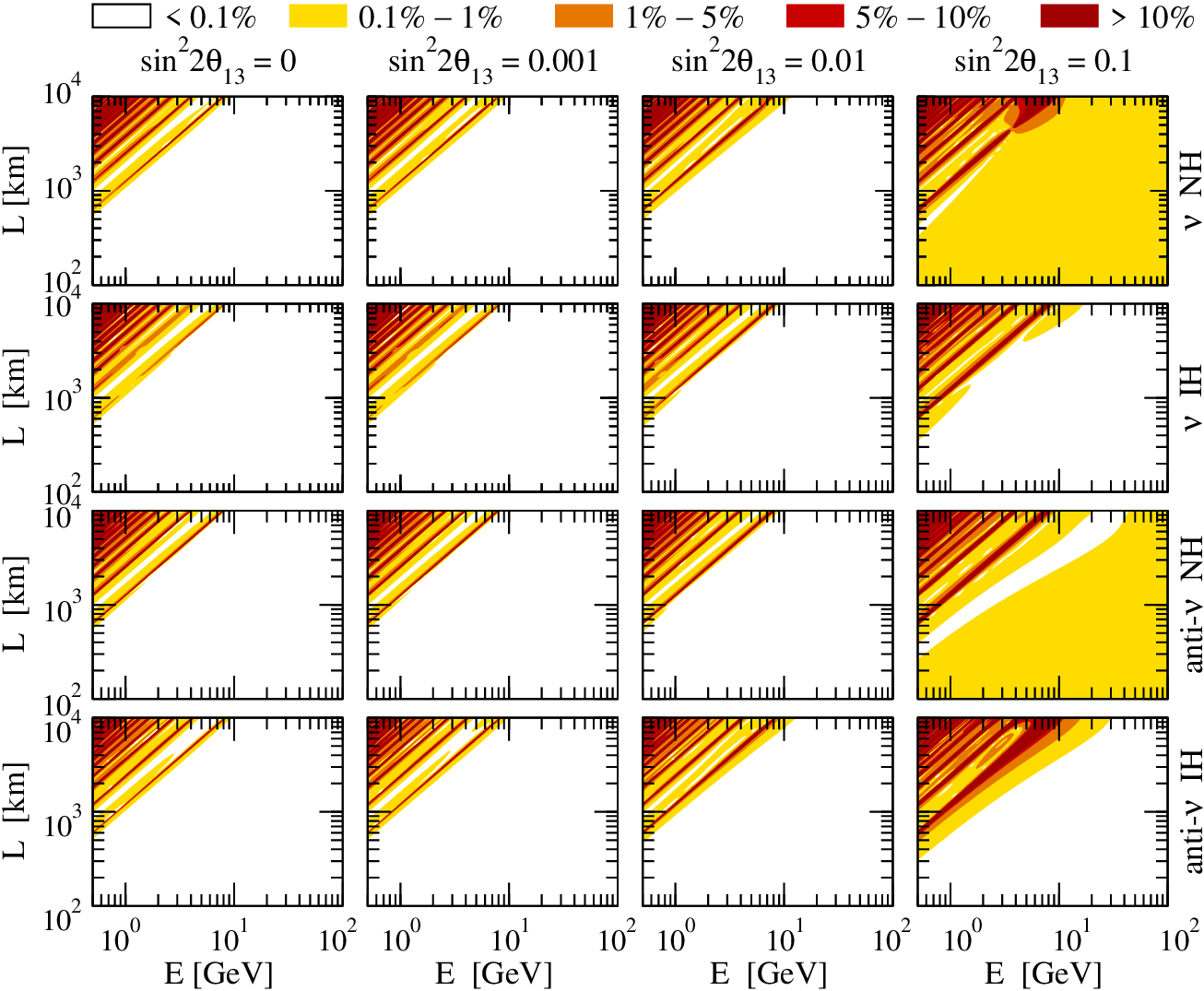}
     \caption{\label{fig:double-Pmutau} Relative errors of the
    double expansion for $P_{\mu\tau}$ for neutrinos (rows 1 and 2), and 
    $P_{\overline{\mu}\overline{\tau}}$ for antineutrinos (rows 3 and 4),
    for normal hierarchy (NH) and inverted hierarchy (IH). The values of the 
    fundamental neutrino parameters used are $|\Delta m_{31}^2|
     = 2 \cdot 10^{-3} \, {\rm eV}^2$, $\alpha = 0.03$, $\theta_{12} =
     33^\circ$, $\theta_{23}= 45^\circ$, $\delta_{\rm CP} = 0$, and
     several values of $\sin^22\theta_{13}$. Matter of constant  
     density $\rho = 3 \, {\rm g/cm^3}$ is assumed.}
\end{figure}

Motivated by the fact that the double expansion is most accurate in a wider 
region of the parameter space than the single expansions, we present
in this subsection some more accuracy tests for it.
Figures~\ref{fig:double-Pemu} and \ref{fig:double-Pmutau} show the
relative errors for $P_{e\mu}$ given in \eq~(\ref{eq:Pem}) and
$P_{\mu\tau}$ given in \eq~(\ref{eq:Pmt}), respectively, for
neutrinos, antineutrinos, and normal and inverted hierarchies. 
We note that the relative accuracy of $P_{e\tau}$ and $1-P_{ee}$ is 
very similar to the one of $P_{e\mu}$ (\fig~\ref{fig:double-Pemu}), and the 
accuracy of $P_{\mu\mu}$ and $P_{\tau\tau}$ is similar to that of 
$P_{\mu\tau}$ (\fig~\ref{fig:double-Pmutau}). 

In the region of $L/E$ defined in \eq~(\ref{eq:atm_region}), the
accuracy of $P_{e\mu}$ is roughly between 1\% and 5\%.  It should be
noted that the absolute errors of the double expansion of $P_{e\mu}$
depend very weakly on the choice of the fundamental neutrino 
parameters within their allowed ranges, and they are always very
small, similar to the first row of plots in
\fig~\ref{fig:compare-abs}. However, the relative errors shown in 
\fig~\ref{fig:double-Pemu} are rather sensitive to the fundamental
parameters values in the region of low $L/E$. The reason is that the
probability $P_{e\mu}$ itself is typically very small in this region,
being as tiny as $10^{-7}$, or even smaller. Hence, the relative error,
being a ratio of two small numbers, is quite sensitive to variations
of the parameters (see, \eg, the second column in
\fig~\ref{fig:double-Pemu}). 
In contrast, $P_{\mu\tau}$ is rather large, because of the oscillations in 
``atmospheric'' channel driven by the large $\Delta m_{31}^2$ and maximal or 
nearly maximal mixing $\sin^22\theta_{23} \simeq 1$. This is also clear from 
\eq~(\ref{eq:Pmt}), since $P_{\mu\tau}$ has a term of zeroth order in both 
$\alpha$ and $s_{13}$, while $P_{e\mu}$ from \eq~(\ref{eq:Pem}) is of second 
order in these parameters. Hence, in the case of $P_{\mu\tau}$, the 
relative errors are of the same order as the absolute errors, which are  
always very small. The errors shown in \fig~\ref{fig:double-Pmutau} are 
smaller than 0.1\% for $\sin^22\theta_{13} \le 0.01$ and smaller than 1\% 
for $\sin^22\theta_{13} \le 0.1$.

\subsection{Probability expansions and relation to experiments}
\label{sec:experiments}

The discussion of the accuracy of the different expansions can be used
to identify the best set of equations for a given neutrino oscillation
experiment. The first question is, however, whether matter effects are
relevant or if the much simpler vacuum probabilities can be used.  The
absolute and relative differences between matter and vacuum
probabilities vanish, as discussed in \Sec~\ref{sec:Pemu}, for
sufficiently low energies or for short enough baselines. The
discussion showed that in experimentally interesting cases matter
effects are almost always relevant for the ranges of $L$ and $E$ that
we considered. If $L/E$ is such that oscillations can be observed,
then for fixed $L/E$ the matter effects are negligible only at very
small $E$ (and consequently small $L$). An example where matter
effects can be ignored at the percent level is given by the reactor
$\bar\nu_e$ disappearance experiments with energies of a few MeV
and baselines up to a few kilometers.

In order to identify the best expansion for a given experiment, one
must take into account that the analyses of real experiments are based
on event rates and not on probabilities. It is useful to distinguish
here between ``point sources'' (accelerator beams, reactors, the Sun,
supernovae, \etc) where the observed neutrinos come from one 
point-like region in space, and ``extended sources'' (\eg, the
atmosphere) where the observed neutrinos come from many different
positions in space. An event rate based analysis of point sources
implies essentially, for a given energy, an additional factor $1/L^2$
in order to account for the reduction of the flux as a function of the
baseline $L$. In addition, the detection cross section is typically
proportional to $E^r$, where $1 \lesssim r \lesssim 2$, depending on
the energy range and on the detection process. In the two-flavor
approximation, when going from the first oscillation maximum at
$\Delta_{2\nu}=\pi/2$ to the $n$th oscillation maximum at
$\Delta_{2\nu}=(n - 1/2) \pi$ this implies that the event rates drop
by a factor $1/(2n-1)^2$ if the baseline is increased, or by a factor
$1/(2n-1)^r$ if the energy is decreased. Hence, \eg, for a fixed
energy the rate in the second (third) maximum is already reduced by a
factor 1/9 (1/25). 
This explains why most current proposals for long baseline oscillation
experiments aim at the first oscillation maximum of the
``atmospheric'' oscillations. The oscillations driven by the ``solar''
frequency are then a sub-leading effect, and in most cases the double
expansion works very well. The overall relative precision of the
double expansion is then typically in the range of a percent, or at worst,
up to a few percent in this case.  However, if $\theta_{13}$ is close
to its current upper limit, the single expansion in $\alpha$ is even
more precise than the double expansion. On the other hand, if
$\theta_{13}$ is tiny, well below the sensitivity limits of any
planned accelerator experiment, the single expansion in $s_{13}$ is
more precise than the double expansion, especially for low
energies. Note, however, that the double expansion has often a
sufficient precision even when it is not the best expansion.

The discussion of point sources does not change much for experiments 
which aim at higher oscillation maxima. The
$1/(2n-1)^2$ flux factor due to the beam divergence and/or the $1/(2n-1)^r$ 
factor due to the energy dependence of the detection cross sections 
reduce the observed event rates and limit all proposals to the first few 
oscillation maxima at best. The double expansion still works quite well for 
such proposals, as long as the sub-leading oscillations governed by the
``solar'' frequency stay in the linear regime, where an expansion in 
$\alpha$ makes sense. This leads to the condition $2n-1 \ll \alpha^{-1}\simeq 
40$, which is fulfilled for the first few maxima. Higher values of $n$ are 
currently not proposed, since the flux and/or cross section would drop by 
a large factor, which would make the detectors and sources unaffordable. For 
largest (smallest) $\theta_{13}$ the single expansion in $\alpha$ ($s_{13}$) 
works again numerically even better then the double expansion, just like in 
the case of the first oscillation maximum. 

The discussion is somewhat different for sources which are not
point-like.  In that case, there is in general no $1/L^2$ suppression
of the flux, and the observed event rates are not dominated by the
first oscillation maximum.  An important example are atmospheric
neutrinos, for which a wide range of baselines and energies
contributes: the energy window of practical interest ranges from a
100~MeV to a few tens of GeV, while the baseline lies between $10$~km
and $10^4$~km. This wide range of energies and baselines makes clear
that none of the discussed expansions covers the full parameter range
at the percent level. In general the double expansion works quite
well for small or moderate $L/E$, while the $s_{13}$ expansion tends
to work best for larger $L/E$ values. At the same time, close to the
atmospheric resonance or if the mixing angle $\theta_{13}$ is near its
current upper bound and for not too large $L/E$, the $\alpha$
expansion is expected to be the most relevant one.

In general, for all types of experiments and for any given values of 
$L$, $E$, and $\theta_{13}$ the most accurate expansion can readily be 
found with the help of \fig~\ref{fig:which_best}.

\section{Summary and conclusions}
\label{sec:conclusions}

In this paper we have presented three different sets of approximate
formulas for the probabilities of neutrino oscillations in matter and
in vacuum. We have shown that in general the probabilities for all
possible oscillation channels for neutrinos as well as for
antineutrinos can be obtained from just two independent probabilities
by using unitarity and a symmetry related to the ``atmospheric''
mixing angle $\theta_{23}$ [see \eq~(\ref{eq:notat})]. One possible
choice for these two probabilities is $P_{e\mu}$ and $P_{\mu\tau}$.
We have derived the expressions for the neutrino oscillation
probabilities in matter of constant density, expanded in terms of the
small parameters $\alpha \equiv \Delta m^2_{21} / \Delta m^2_{31}$,
$s_{13}$ or in both of them. Below we summarize the main features of
these expansions:
\begin{itemize}
\item \textbf{Double expansion up to second order in
\boldmath$\alpha$ and \boldmath$s_{13}$}
[\Sec~\ref{sec:double_const}, \eqs~(\ref{eq:Pee})--(\ref{eq:Pmt})]

The neutrino oscillation probabilities are expanded in both small
parameters up to second order. In general, these expressions are valid
for $ \alpha \Delta = \Delta m_{21}^2 L /(4E) \ll 1$ or $L/E \ll 10^4
\, \mathrm{km}/\mathrm{GeV}$, \ie, if neutrino oscillations due to the
solar mass squared difference $\Delta m^2_{21}$ are not important.
The accuracy of the approximation is good for a wide range of the
parameters. Typically, the relative errors of $P_{e\mu}$, $P_{e\tau}$,
and $1-P_{ee}$ are between 1\% and 5\%, the relative errors of
$P_{\mu\mu}$, $P_{\mu\tau}$, and $P_{\tau\tau}$ and the absolute
errors for all probabilities are of the order 0.1\%.

\item \textbf{Single expansion up to first order in \boldmath$\alpha$}
[\Sec~\ref{sec:alpha_const}, 
\eqs~(\ref{eq:Pee0_alpha})--(\ref{eq:Pmt1_alpha})]

The neutrino oscillation probabilities are expanded in $\alpha$, but the 
exact dependence on $s_{13}$ is retained. Like in the case of the double 
expansion, these formulas are valid in the region where the oscillations 
driven by the solar mass squared difference are not important: $\alpha 
\Delta = \Delta m_{21}^2 L /(4E) \ll 1$, or $L/E \ll 10^4 \, \mathrm{km} / 
\mathrm{GeV}$. The accuracy is better than the one of the double 
expansion for values of $s_{13} \gtrsim 0.1$ close the current upper bound, 
or if the atmospheric resonance is important.  For instance, for a matter 
density $\rho =3\, {\rm g/cm^3}$ and $\Delta m^2_{31} = 2 \cdot 10^{-3} \, 
{\rm eV}^2$ this is the case for $E \sim 10$~GeV and $L \gtrsim 5500$~km.

\item \textbf{Single expansion up to first order in \boldmath$s_{13}$}
[\Sec~\ref{sec:s13_const}, 
\eqs~(\ref{eq:Pee0_theta})--(\ref{eq:Pmt1_theta})]

The neutrino oscillation probabilities are expanded in $s_{13}$, but
the exact dependence on $\alpha$---and therefore on $\Delta
m^2_{21}$---is retained.  Hence, these expressions are useful in the
region where the oscillations due to the solar mass squared difference
are relevant: $\alpha \Delta = \Delta m_{21}^2 L / (4E) \gtrsim 1$, or
$L/E \gtrsim 10^4 \, \mathrm{km} / \mathrm{GeV}$. In particular, this
is the case for low energies $E \sim 0.1$~GeV and $L\gtrsim 10^3$
km. For values of $L/E$ outside the ``solar'' regime this type of
expansion is only useful for very small values of $s_{13} \lesssim
10^{-4} \div 10^{-3}$.

\item \textbf{Vacuum limit of the expansions}
[\Secs~\ref{sec:double_vac}, \ref{sec:alpha_vac}, \ref{sec:s13_vac}]

The vacuum limit for each type of expansion is valid if, in addition to 
the requirements ensuring the validity of a given type of expansion,  
matter effects can be neglected (see discussion in \Sec~\ref{sec:Pemu}). 
This usually implies low energies or short baselines. We find that in many 
realistic cases matter effects are of the order of a few percent. Hence, if an 
accuracy at that level is required, one should employ the formulas for  
neutrino oscillations in matter.
\end{itemize}

In conclusion, we have presented a collection of formulas for
three-flavor neutrino oscillation probabilities by deriving expansions
in small parameters. We have performed a detailed analysis of the
accuracy of these expansions and determined the parameter regions
where they are most accurate. The expansions of the neutrino
oscillation probabilities in matter of constant density are useful for
the analytical understanding of the physics of future neutrino
oscillation experiments. Furthermore, we have also presented expansion
formulas for the neutrino oscillation probabilities in arbitrary
matter density profiles (see \App~\ref{app:pert_evol}), which can be
applied to a large class of problems.


\subsection*{Acknowledgments}

We would like to thank Martin Freund, H{\aa}kan Snellman, and Walter
Winter for useful discussions and comments. T.S.~thanks the KTH for
hospitality and financial support for a research visit.  R.J.~and
T.O.~would like to thank the TUM for the warm hospitality during their
research visits as well as for the financial support.  E.A.~was
supported by the sabbatical grant SAB2002-0069 of the Spanish Ministry
of Education, Culture, and Sports, the RTN grant HPRN-CT-2000-00148 of
the European Commission, the ESF Neutrino Astrophysics Network and
MCyT grant BFM2002-00345. M.L.~and T.S.~were supported by the
``Sonderforschungsbereich 375 f{\"u}r Astro-Teilchenphysik der
Deutschen For\-schungs\-ge\-mein\-schaft'', and T.O.~was supported by
the Swedish Research Council (Vetenskapsr{\aa}det), Contract
Nos.~621-2001-1611, 621-2002-3577, the G{\"o}ran Gustafsson
Foundation, and the Magnus Bergvall Foundation.


\begin{appendix}
\renewcommand{\theequation}{\thesection\arabic{equation}}

\setcounter{equation}{0}
\section{Hamiltonian diagonalization approach} 
\label{app:diagonalization}

In this appendix, we present the details of the approximate diagonalization 
of the effective Hamiltonian of the neutrino system in the case of matter 
of constant density. 

\subsection{Diagonalizing the Hamiltonian perturbatively}
\label{app:perturbation}

In order to derive the double expansions given in
\Sec~\ref{sec:double}, we write the Hamiltonian of \eq~(\ref{eq:ham1})
as
\begin{equation}
H \simeq \frac{\Delta m_{31}^2}{2E} 
O_{23} U_\delta \, M \, U_\delta^\dagger O_{23}^T \,,
\label{eq:defM}
\end{equation}
where $M \equiv O_{13}O_{12} \, \mathrm{diag}\left(0, \alpha, 1
\right) O_{12}^T O_{13}^T + \mathrm{diag}\left(A, 0, 0 \right)$, and
the matrices $O_{ij}$ and $U_\delta$ have been defined after
\eq~(\ref{eq:U}). The matrix $M$ can be explicitly written as $M=
(\Delta m_{31}^2/2E)^{-1}H'$ by setting $\delta_\mathrm{CP} = 0$ in
$H'$, which is given in \eq~(\ref{eq:ham2}) below.
First, we diagonalize the matrix $M$ by $M = W \hat M W^\dagger$ with
$\hat M = \mathrm{diag}(\lambda_1,\lambda_2,\lambda_3)$ and $W$ being
a unitary diagonalizing matrix. This diagonalization is performed by
using perturbation theory up to second order in the small parameters
$\alpha$ and $s_{13}$, \ie, we write $M = M^{(0)} + M^{(1)} +
M^{(2)}$, where $M^{(1)}$ ($M^{(2)}$) contains all terms of first
(second) order in $\alpha$ and $s_{13}$. One finds
\begin{align}
M^{(0)} &= \mathrm{diag}(A,0,1) = \mathrm{diag}(\lambda_1^{(0)},
\lambda_2^{(0)}, \lambda_3^{(0)}) \,, \\
M^{(1)} &= 
  \left(\begin{array}{ccc}
     \alpha s_{12}^2  &  \alpha s_{12} c_{12}  & s_{13} \\
     \alpha s_{12} c_{12}  &  \alpha c_{12}^2  & 0  \\
     s_{13} & 0 & 0
  \end{array}\right) \,, \\
M^{(2)} &= 
  \left(\begin{array}{ccc}
     s_{13}^2 & 0 & -\alpha s_{13} s_{12}^2  \\
     0 & 0 & -\alpha s_{13} s_{12} c_{12} \\
     -\alpha s_{13} s_{12}^2  & -\alpha s_{13} s_{12} c_{12} & -s_{13}^2
  \end{array}\right) \,.
\end{align}
For the eigenvectors we write $v_i = v_i^{(0)} + v_i^{(1)} + v_i^{(2)}$, and
since $M^{(0)}$ is diagonal at zeroth order, we have $v_i^{(0)} = e_i$. Then, 
the first and second order corrections to the eigenvalues are given by
\begin{align}
\lambda_i^{(1)} &= M^{(1)}_{ii}  \,, \\
\lambda_i^{(2)} &= M^{(2)}_{ii} 
+ \sum_{j \neq i} 
\frac{\left( M^{(1)}_{ii} \right)^2}{\lambda_i^{(0)} - \lambda_j^{(0)}}
\,,
\end{align}
and the corrections to the eigenvectors are calculated by
\begin{align}
v_i^{(1)} &= \sum_{j \neq i} 
\frac{ M^{(1)}_{ij} }{\lambda_i^{(0)} - \lambda_j^{(0)}} \: e_j
\,, \\
v_i^{(2)} &= \sum_{j \neq i}
\frac{ 1 }{\lambda_i^{(0)} - \lambda_j^{(0)}} 
\left[ M^{(2)}_{ij} + \left( M^{(1)} v_i^{(1)} \right)_j
- \lambda^{(1)}_i \left( v_i^{(1)} \right)_j \right] e_j \,.
\end{align}

The mixing matrix in matter is given by $U' = O_{23} U_\delta W$ with
$W = (v_1, v_2, v_3)$, and the eigenvalues of the Hamiltonian [see
\eqs~(\ref{eq:EV1})--(\ref{eq:EV3})] are obtained as $E_i = [\Delta
m_{31}^2 / (2E)] \lambda_i$ ($i=1,2,3$). 

In our paper, we do not in general order the eigenvalues according to their 
magnitude.  Such an ordering would create problems as one would have to 
re-label the eigenvalues  upon passing through each of the two MSW resonances. 
The ordering is actually unimportant if one is careful to assign the correct 
eigenvector to each eigenvalue. 

We also reiterate the point discussed at the end of
\Sec~\ref{sec:double_const}: despite the fact that in the case of the
double expansion the eigenvalues as well as certain entries of the
leptonic mixing matrix in matter are divergent at $A\to 0$ and $A\to
1$, the neutrino oscillation probabilities are finite in these
limits. In particular, the correct vacuum probabilities are recovered
in the limit $A\to 0$.  The mentioned divergences are of no concern to
us, since we are interested in oscillation probabilities rather than
in eigenvalues or the mapping between the mixing in matter and in vacuum,
as was the case, \eg, in \Ref~\cite{Freund:2001pn}.

\subsection{The Cayley-Hamilton formalism}
\label{app:cayley-hamilton}

In order to derive a series expansion for the evolution matrix $S$ in
small parameters, we can use the following formula from
\Refs~\cite{Ohlsson:1999xb,Ohlsson:2001vp}:
\begin{align} \label{eq:cayley-hamilton}
S(t,t_0) &= {\rm e}^{-\imag H L}  \nonumber \\
&= \frac{{\rm e}^{-\imag E_1 L}}{(E_1-E_2)(E_1-E_3)} \,
   \big[E_2E_3 \mathbbm{1}_3-(E_2+E_3)H + H^2 \big]
\nonumber \\
&+\frac{{\rm e}^{-\imag E_2 L}}{(E_2-E_1)(E_2-E_3)} \,
\big[E_1E_3 \mathbbm{1}_3-(E_1+E_3)H+H^2 \big]
\nonumber \\
&+\frac{{\rm e}^{-\imag E_3 L}}{(E_3-E_1)(E_3-E_2)}  \,
\big[E_1E_2 \mathbbm{1}_3-(E_1+E_2)H+H^2 \big] \,, 
\end{align}
where $H$ is the Hamiltonian in the flavor basis as given in
\eq~(\ref{eq:ham1}) and $E_i$ ($i=1,2,3$) are the eigenvalues of this
Hamiltonian. Inserting series expansions of the eigenvalues [see, \eg,
\eqs~(\ref{eq:alpha_EV_1})--(\ref{eq:alpha_EV_3}) and
(\ref{eq:theta_EV_1})--(\ref{eq:theta_EV_3})] and the Hamiltonian $H$
into \eq~(\ref{eq:cayley-hamilton}), it is
straightforward to obtain an expression for $S_{\alpha\beta}$, and
subsequently, for the neutrino oscillation probabilities $P_{\alpha\beta} =
|S_{\beta\alpha}|^2$.

\setcounter{equation}{0}
\section{Perturbative expansion of the evolution 
equation for arbitrary matter density profiles}
\label{app:pert_evol}

In this appendix, we present the details of the perturbative expansion of 
the neutrino evolution equation in the case of matter with an arbitrary 
density profile. We also give the formulas relevant for the particular 
case of matter of constant density. 

Consider the neutrino evolution matrix $S(t, t_0)$ defined in 
\eq~(\ref{eq:S}). It is convenient to rotate it by the angle $\theta_{23}$ 
according to \eq~(\ref{eq:SS'}). The evolution matrix in the rotated basis 
$S'(t,\,t_0)$ satisfies the evolution equation ${\rm i}({\rm d}/{\rm 
d}t)S'(t, t_0)=H'(t) S'(t,t_0)$ with the initial condition $S'(t_0, t_0)=
\mathbbm{1}$ and the Hamiltonian $H'(t)$ given by \cite{Akhmedov:2001kd}
\begin{equation}
H'(t) = 
\left( \begin{array}{ccc} s_{12}^2 c_{13}^2 \Delta_{21} + s_{13}^2
\Delta_{31} + V(t) & s_{12} c_{12} c_{13} \Delta_{21} & s_{13} c_{13} 
\left(\Delta_{31} - s_{12}^2 \Delta_{21} \right) {\rm e}^{-{\rm i}
\delta_{CP}} \\
s_{12} c_{12} c_{13} \Delta_{21} & c_{12}^2 \Delta_{21} & - s_{12} c_{12} 
s_{13} {\rm e}^{-{\rm i} \delta_{CP}} \Delta_{21} \\ s_{13} c_{13} \left( 
\Delta_{31} - s_{12}^2 \Delta_{21}\right) {\rm e}^{{\rm i} \delta_{CP}} & 
- s_{12} c_{12} s_{13} {\rm e}^{{\rm i}\delta_{CP}} \Delta_{21} &
c_{13}^2 \Delta_{31} + s_{12}^2 s_{13}^2 \Delta_{21} 
\end{array}
\right).
\label{eq:ham2}
\end{equation}
Here
\begin{equation}
\Delta_{21}\equiv \frac{\Delta m_{21}^2}{2E}\,,\qquad
\Delta_{31}\equiv \frac{\Delta m_{31}^2}{2E}\,.
\end{equation}
It is important to notice that $H'(t)$ [and hence, $S'(t, t_0)$] does
not depend on the mixing angle $\theta_{23}$. This is a consequence of
the specific parameterization of the leptonic mixing matrix $U$ in
\eq~(\ref{eq:U}), in which the matrix $O_{23}$ is the leftmost one,
and the fact that the matrix of the matter-induced potentials in the
effective Hamiltonian $H(t)$ commutes with $O_{23}$. Once the matrix
$S'(t, t_0)$ is found, one can use \eq~(\ref{eq:SS'}) to rotate back
to the original flavor basis. 

It is convenient to decompose the Hamiltonian $H'(t)$ as $H'(t) = H_0'(t) + 
H_I'$, where $H'_0(t)$ is of zeroth order in some expansion parameter and 
$H_I'$ is the remainder. Accordingly, the evolution matrix can be 
written as $S'(t, t_0)=S_0'(t, t_0) S_1'(t, t_0)$, where $S_0(t, t_0)$ 
satisfies
\begin{equation}
{\rm i}\frac{{\rm d}}{{\rm d}t}S_0'(t,t_0)=H_0'(t) S_0'(t,t_0)\,,\qquad 
S_0'(t_0,t_0)=\mathbbm{1}\,.
\label{SS2}
\end{equation}
Then the matrix $S_1'(t, t_0)$ satisfies  
\begin{equation}
{\rm i}\frac{{\rm d}}{{\rm d}t}S_1'(t,t_0) = [S_0'(t,t_0)^{-1} H_I' 
S_0'(t,t_0)]S_1'(t,t_0)\,,
\qquad S_1'(t_0,t_0)=\mathbbm{1}\,.
\label{S1n}
\end{equation}
Up to now, no approximations have been made. Next, we shall find the 
evolution matrix perturbatively. Let $H_1'$ be the part of $H_I'$ that 
is of the first order in the chosen expansion parameter. Then, to first 
order in this parameter, one finds
\begin{equation}
S'(t,t_0) \simeq S_0'(t,t_0)-{\rm i}S_0'(t,t_0)\int_{t_0}^t
[S_0'(t',t_0)^{-1} H_1' S_0'(t',t_0)]\,{\rm d}t'\,, 
\label{Sprime}
\end{equation}
from which, upon rotation back to the original flavor basis, the evolution 
matrix $S(t, t_0)$ is obtained. We shall now consider the application of 
the above formalism to the cases when the expansion parameter is either 
$\alpha$ or $s_{13}$.  

\subsection{Expansion up to first order in $\mbox{\boldmath$\alpha$}$}
\label{app:arbitrary_alpha}

In this case, the zeroth-order Hamiltonian $H_0'(t)$ is obtained from 
\eq~(\ref{eq:ham2}) by taking the limit $\alpha \to 0$ (\ie,
$\Delta_{21} \to 0$). 
The zeroth-order evolution matrix $S_0'(t, t_0)$ can then be written as 
\cite{Akhmedov:1998xq}
\begin{equation}
S_0'(t,t_0)=\left(
\begin{array}{ccc}
u(t,t_0) & 0 & v(t,t_0)\, {\rm e}^{-{\rm i}\delta_{\rm CP}}\\
0 & f(t,t_0) & 0 \\
-v^*(t,t_0)\, {\rm e}^{{\rm i}\delta_{\rm CP}} & 0 & u^*(t,t_0)
\end{array} \right)\,,
\end{equation}
where $u(t,t_0)$ and $v(t,t_0)$ are to be found from the solution to the 
two-flavor problem governed by the mass squared difference $\Delta m_{31}^2$, 
mixing angle $\theta_{13}$, and matter potential $V(t)$, and 
\begin{equation}
f(t,t_0)=\exp\left[{{\rm i}\!\int_{t_0}^t \tilde{\Delta}(t')\,{\rm 
d}t'}\right]\,,
\qquad \tilde{\Delta}(t)\equiv \frac{1}{2}\left[\frac{\Delta m_{31}^2}
{2E}+V(t)\right]\,. 
\end{equation}
The parameters $u$ and $v$ satisfy $|u|^2+|v|^2=1$. The factor ${\rm
  e}^{-{\rm i} 
\delta_{\rm CP}}$ was pulled out from $v$ for convenience (this will simplify 
some of the formulas given below). 

The neutrino oscillation probabilities to first order in $\alpha$ in
the case of matter with an arbitrary density profile can be written as 
\begin{align}
P_{ee} &= |u|^2 - 2 \, \frac{\alpha\Delta}{L} \, s_{12}^2 \, {\rm Im} [u v^* 
\,(2\cos 2\theta_{13} \, I_1 + \sin 2\theta_{13} \, I_2 )]\,,
\label{eq:Pee_alpha_arb}\\
P_{e\mu} &= s_{23}^2 |v|^2 + 2 \, \frac{\alpha\Delta}{L} \, s_{12}^2
\, s_{23}^2 \, {\rm Im}  [u v^* \,(2\cos 2\theta_{13}\, I_1 +\sin
  2\theta_{13}\, I_2)]
+\frac{\alpha\Delta}{L}\,\sin 2\theta_{12}\sin 2\theta_{23} \nonumber\\
&\times \left\{ \cos\delta_{\rm CP} \, {\rm Im} [v^* f^*(c_{13} I_3+s_{13} 
I_4)] +
\sin\delta_{\rm CP} \, {\rm Re}
[v^* f^*(c_{13} I_3+s_{13} I_4)]\right\}\,, \\
P_{\mu\tau} &=
\frac{1}{4} \sin^2 2\theta_{23}\,[1+|u|^2-2 {\rm Re}(u f)]
\nonumber\\
&-\frac{\alpha\Delta}{L}\,\sin^2 2\theta_{23}
\left\{s_{12}^2\,{\rm Im} \left[(u-f^*) v^* \left(\cos
  2\theta_{13}\,I_1+\frac{1}{2}\sin 2\theta_{13}\,I_2 \right)
  \right]\right. \nonumber\\
&+\left. \left[c_{12}^2 L-s_{12}^2 \left(s_{13}^2\,I_5+c_{13}^2\,I_6 -
  \frac{1}{2} \sin 2\theta_{13} \,I_7\right) \right]\, {\rm Im} (u f) \right\}
\nonumber\\
&-\frac{\alpha\Delta}{L}\, \sin 2\theta_{12}\,\sin
2\theta_{23} \Big\{\cos\delta_{\rm CP} \big[ c_{23}^2 \,{\rm Im} [(u-f^*) v^*
  (c_{13}\,I_3+s_{13}\,I_4)] \nonumber\\
&+ {\rm Im} [(c_{23}^2 |u|^2-s_{23}^2-\cos 2\theta_{23} \, u f)
    \,(c_{13}\,I_9-s_{13}\,I_8)]\big] \nonumber\\
&+\sin\delta_{\rm CP} \big[ c_{23}^2 \,{\rm Re}[(u-f^*) v^*
    (c_{13}\,I_3+s_{13}\,I_4)] \nonumber\\
&-{\rm Re} [(c_{23}^2 |u|^2-s_{23}^2 - \cos
  2\theta_{23} \, u f) \, (c_{13}\,I_9-s_{13}\,I_8)]\big]\Big\}\,,
\label{eq:Pmt_alpha_arb}
\end{align}
where the quantities $I_1,I_2,\ldots,I_9$ are given by
\begin{align}
& I_1 \equiv \int_{t_0}^t u^*(t',t_0) v(t',t_0) \, \mathrm{d}t'\,, 
\qquad 
I_2 \equiv \int_{t_0}^t [v(t',t_0)^2-u^*(t',t_0)^2] \, \mathrm{d}t'\,, 
\nonumber \\
& I_3 \equiv \int_{t_0}^t u^*(t',t_0) f(t',t_0) \, \mathrm{d}t'\,, 
\qquad 
I_4 \equiv \int_{t_0}^t v(t',t_0) f(t',t_0) \, \mathrm{d}t'\,, 
 \nonumber \\
& I_5 \equiv \int_{t_0}^t |u(t',t_0)|^2 \, \mathrm{d}t'\,,\qquad \qquad
~~I_6 \equiv \int_{t_0}^t |v(t',t_0)|^2 \, \mathrm{d}t'\,, 
\nonumber \\
& I_7 \equiv \int_{t_0}^t [u(t',t_0)v(t',t_0)+u^*(t',t_0)v^*(t',t_0) ]
\, \mathrm{d}t'\,, \nonumber \\
& I_8 \equiv \int_{t_0}^t u^*(t',t_0) f^*(t',t_0) \, \mathrm{d}t'\,, 
\qquad  
I_9 \equiv \int_{t_0}^t v(t',t_0) f^*(t',t_0) \, \mathrm{d}t'\,.
\nonumber
\end{align}
Note that the integrals $I_5$, $I_6$, and $I_7$ are real, $I_5+I_6=t-t_0$,  
and the integrals $I_8$ and $I_9$ can be obtained from $I_3$ and $I_4$, 
respectively, by substituting $f \to f^*$. 

In the case of matter of constant density, the parameters $u$, $v$, and
$f$ are given by
\begin{align}
u(t,0) &= \cos \frac{C_{13}\Delta \,t}{L} + {\rm i} \frac{\cos
  2\theta_{13} - A}{C_{13}} \sin \frac{C_{13}\Delta \,t}{L} \,, \\
v(t,0) &= - {\rm i} \frac{\sin 2\theta_{13}}{C_{13}} \sin
  \frac{C_{13}\Delta \,t}{L} \,, \\
f(t,0) &= \exp\left[{{\rm i} \frac{(1+A)\Delta \,}{L} t}\right] \,.
\end{align}
Inserting these expressions into the integrals $I_1,I_2,\ldots,I_9$ and the 
obtained results into \eqs~(\ref{eq:Pee_alpha_arb})--(\ref{eq:Pmt_alpha_arb}) 
leads to \eqs~(\ref{eq:Pee0_alpha})--(\ref{eq:Pmt1_alpha}).

\subsection{Expansion up to first order in $\mbox{\boldmath$s_{13}$}$}
\label{app:arbitrary_s13}

This case was studied in detail in \Ref~\cite{Akhmedov:2001kd}
(\App~A), which we closely follow here. The zeroth-order Hamiltonian
$H_0'(t)$ is obtained from \eq~(\ref{eq:ham2}) by taking the limit
$\theta_{13} \to 0$.  The zeroth-order evolution matrix $S_0'(t, t_0)$
can be written as
\begin{equation}
S_0'(t,t_0)=\left(\begin{array}{ccc}
x(t,t_0)   & y(t,t_0)    & 0 \\
-y^*(t,t_0)  &  x^*(t,t_0) & 0 \\
0 & 0 & g(t,t_0)
\end{array}\right)\,, 
\end{equation}
where $x(t,t_0)$ and $y(t,t_0)$ are to be found from the solution to the 
two-flavor problem governed by the mass squared difference $\Delta m_{21}^2$, 
mixing angle $\theta_{12}$, and matter potential $V(t)$, and 
\begin{equation}
g(t,t_0)=\exp\left[{-{\rm i}\!\int_{t_0}^t \hat\Delta(t')\,{\rm 
d}t'}\right]\,,
\qquad \hat\Delta(t)\equiv \frac{\Delta m_{31}^2}{2E}-\frac{1}{2}
\left[\frac{\Delta m_{21}^2}{2E}+V(t)\right]\,. 
\end{equation}
The parameters $x$ and $y$ satisfy $|x|^2+|y|^2=1$. 

The neutrino oscillation probabilities to first order in $s_{13}$ in
the case of matter with an arbitrary density profile can be written as 
\begin{align} \label{eq:Pee_s13_arb}
P_{ee} &= |x|^2\,, \\
P_{e\mu} &= c_{23}^2\,|y|^2-\sin 2\theta_{23}\,{\rm Im} (y g C)\,, \\
P_{\mu\tau} &= s_{23}^2 c_{23}^2\,|x-g^*|^2+\sin
2\theta_{23}\,{\rm Im} [(x-g^*)g(s_{23}^2 B-c_{23}^2 D)]\,,
\label{eq:Pmt_s13_arb}
\end{align}
where the quantities $B$,$C$, and $D$ are defined as \cite{Jacobson:2003wc}
\begin{align} \label{eq:B}
B \equiv B(t,t_0) &= a I_{y^*,t}(t,t_0) + b I_{x,t}(t,t_0)\,, \\
C \equiv C(t,t_0) &= a^* I^*_{x^*,t_0}(t,t_0) - b^* I^*_{y,t_0}(t,t_0)\,, \\
D \equiv D(t,t_0) &= a^* I^*_{y^*,t_0}(t,t_0) + b^* I^*_{x,t_0}(t,t_0)\,.
\end{align}
Here the integral $I_{\varphi,s}$ is defined as 
\begin{equation}
I_{\varphi,s}(t,t_0) = \int_{t_0}^t \varphi(t',s) g(t',s) \,
\mathrm{d} t' \,,
\label{eq:I}
\end{equation}
and the quantities $a$ and $b$ are given by
\begin{align}
a &= \frac{\Delta m_{31}^2}{2E} s_{13} (1-s_{12}^2 \alpha) {\rm e}^{-{\rm i}
  \delta_{\rm CP}}\,, \\
b &= -\frac{\Delta m_{21}^2}{2E} s_{13} s_{12} c_{12} {\rm e}^{-{\rm i}
  \delta_{\rm CP}}\,.
\end{align}

In the case of matter of constant density, the parameters $x$, $y$, and
$g$ are given by
\begin{align}
x(t,0) &= \cos \frac{\alpha C_{12}\Delta \,t}{L} + {\rm i} \frac{\cos
  2\theta_{12} - A/\alpha}{C_{12}} \sin \frac{\alpha C_{12}\Delta \,t}{L} 
\,, \\
y(t,0) &= - {\rm i} \frac{\sin 2\theta_{12}}{C_{12}} \sin
  \frac{\alpha C_{12}\Delta \,t}{L} \,, \\
g(t,0) &= \exp\left[{{\rm i} \frac{(A+\alpha-2)\Delta \,}{L} t}\right] \,.
\end{align}
Inserting these expressions into \eqs~(\ref{eq:B}) -- (\ref{eq:I}) and the 
obtained results into \eqs~(\ref{eq:Pee_s13_arb})--(\ref{eq:Pmt_s13_arb})
leads to \eqs~(\ref{eq:Pee0_theta})--(\ref{eq:Pmt1_theta}).

\end{appendix}



\begin{thebibliography}{10}

\bibitem{Fukuda:1998mi}
Super-Kamiokande Collaboration, Y. Fukuda et~al.,
\newblock Phys. Rev. Lett. 81 (1998) 1562, hep-ex/9807003;
%
\newblock Phys. Rev. Lett. 82 (1999) 2644, hep-ex/9812014.

\bibitem{SK}
Super-Kamiokande Collaboration, Y. Hayato,
\newblock talk at the HEP2003 conference (Aachen, Germany, 2003), {\tt
  http://eps2003.physik.rwth-aachen.de}.

\bibitem{Ambrosio:2003yz}
MACRO Collaboration, M. Ambrosio et~al.,
\newblock Phys. Lett. B566 (2003) 35, hep-ex/0304037.

\bibitem{Ahn:2002up}
K2K Collaboration, M.H. Ahn et~al.,
\newblock Phys. Rev. Lett. 90 (2003) 041801, hep-ex/0212007.

\bibitem{solar}
B.T. Cleveland et~al.,
\newblock Astrophys. J. 496 (1998) 505;
%
SAGE, J.N. Abdurashitov et~al.,
\newblock J. Exp. Theor. Phys. 95 (2002) 181;
%
GALLEX Collaboration, W. Hampel et~al.,
\newblock Phys. Lett. B447 (1999) 127;
%
GNO Collaboration, M. Altmann et~al.,
\newblock Phys. Lett. B490 (2000) 16, hep-ex/0006034;
%
Super-Kamiokande Collaboration, S. Fukuda et~al.,
\newblock Phys. Lett. B539 (2002) 179, hep-ex/0205075;
%
SNO Collaboration, Q.R. Ahmad et~al.,
\newblock Phys. Rev. Lett. 89 (2002) 011301, nucl-ex/0204008;
%
S.N. Ahmed et~al.,
\newblock nucl-ex/0309004.

\bibitem{Eguchi:2002dm}
KamLAND Collaboration, K. Eguchi et~al.,
\newblock Phys. Rev. Lett. 90 (2003) 021802, hep-ex/0212021.

\bibitem{Apollonio:1999ae}
CHOOZ Collaboration, M. Apollonio et~al.,
\newblock Phys. Lett. B466 (1999) 415, hep-ex/9907037;
%
\newblock Eur. Phys. J. C27 (2003) 331, hep-ex/0301017.

\bibitem{Bilenky:1987st}
S.M. Bilenky,
\newblock Fiz. Elem. Chast. Atom. Yadra 18 (1987) 449;
%
S.M. Bilenky and S.T. Petcov,
\newblock Rev. Mod. Phys. 59 (1987) 671.

\bibitem{Barger:1980tf}
V.D. Barger et~al.,
\newblock Phys. Rev. D22 (1980) 2718.

\bibitem{Kim:1987vg}
C.W. Kim and W.K. Sze,
\newblock Phys. Rev. D35 (1987) 1404.

\bibitem{Zaglauer:1988gz}
H.W. Zaglauer and K.H. Schwarzer,
\newblock Z. Phys. C40 (1988) 273.

\bibitem{Ohlsson:1999xb}
T. Ohlsson and H. Snellman,
\newblock J. Math. Phys. 41 (2000) 2768, hep-ph/9910546,
\newblock 42 (2001) 2345(E);
%
\newblock Phys. Lett. B474 (2000) 153, hep-ph/9912295,
\newblock 480 (2000) 419(E).

\bibitem{Xing:2000gg}
Z.z. Xing,
\newblock Phys. Lett. B487 (2000) 327, hep-ph/0002246.

\bibitem{Ohlsson:2001vp}
T. Ohlsson,
\newblock Phys. Scripta T93 (2001) 18.

\bibitem{Kimura:2002hb}
K. Kimura, A. Takamura and H. Yokomakura,
\newblock Phys. Lett. B537 (2002) 86, hep-ph/0203099;
%
\newblock Phys. Rev. D66 (2002) 073005, hep-ph/0205295.

\bibitem{Harrison:2003fi}
P.F. Harrison, W.G. Scott and T.J. Weiler,
\newblock Phys. Lett. B565 (2003) 159, hep-ph/0305175.

\bibitem{Lehmann:2000ey}
H. Lehmann, P. Osland and T.T. Wu,
\newblock Commun. Math. Phys. 219 (2001) 77, hep-ph/0006213.

\bibitem{Osland:1999et}
P. Osland and T.T. Wu,
\newblock Phys. Rev. D62 (2000) 013008, hep-ph/9912540.

\bibitem{Kuo:1986sk}
T.K. Kuo and J. Pantaleone,
\newblock Phys. Rev. Lett. 57 (1986) 1805.

\bibitem{Smirnov:1987mk}
A.Y. Smirnov,
\newblock Yad. Fiz. 46 (1987) 1152.

\bibitem{Kuo:1987zx}
T.K. Kuo and J. Pantaleone,
\newblock Phys. Rev. D35 (1987) 3432.

\bibitem{Petcov:1987qg}
S.T. Petcov and S. Toshev,
\newblock Phys. Lett. B187 (1987) 120.

\bibitem{Ohlsson:2001et}
T. Ohlsson and H. Snellman,
\newblock Eur. Phys. J. C20 (2001) 507, hep-ph/0103252.

\bibitem{D'Olivo:1996nk}
J.C. D'Olivo and J.A. Oteo,
\newblock Phys. Rev. D54 (1996) 1187.

\bibitem{MSW}
L.~Wolfenstein,
\newblock Phys. Rev. D17 (1978) 2369;
%
S.P. Mikheev and A.Y. Smirnov,
\newblock Sov. J. Nucl. Phys. 42 (1985) 913.

\bibitem{Akhmedov:1998xq}
E.K. Akhmedov et~al.,
\newblock Nucl. Phys. B542 (1999) 3, hep-ph/9808270.

\bibitem{Peres:1999yi}
O.L.G. Peres and A.Y. Smirnov,
\newblock Phys. Lett. B456 (1999) 204, hep-ph/9902312.

\bibitem{Akhmedov:2001kd}
E.K. Akhmedov et~al.,
\newblock Nucl. Phys. B608 (2001) 394, hep-ph/0105029.

\bibitem{Peres:2002ri}
O.L.G. Peres and A.Y. Smirnov,
\newblock Nucl. Phys. Proc. Suppl. 110 (2002) 355, hep-ph/0201069;
%
\newblock hep-ph/0309312.

\bibitem{Yasuda:1999uv}
O. Yasuda,
\newblock Acta Phys. Polon. B30 (1999) 3089, hep-ph/9910428.

\bibitem{Freund:1999gy}
M. Freund et~al.,
\newblock Nucl. Phys. B578 (2000) 27, hep-ph/9912457.

\bibitem{Mocioiu:2001jy}
I. Mocioiu and R. Shrock,
\newblock JHEP 11 (2001) 050, hep-ph/0106139.

\bibitem{Arafune:1997hd}
J. Arafune, M. Koike and J. Sato,
\newblock Phys. Rev. D56 (1997) 3093, hep-ph/9703351.

\bibitem{Brahmachari:2003bk}
B. Brahmachari, S. Choubey and P. Roy,
\newblock Nucl. Phys. B671 (2003) 483, hep-ph/0303078.

\bibitem{Cervera:2000kp}
A. Cervera et~al.,
\newblock Nucl. Phys. B579 (2000) 17, hep-ph/0002108,
\newblock 593 (2001) 731(E).

\bibitem{Freund:2001pn}
M. Freund,
\newblock Phys. Rev. D64 (2001) 053003, hep-ph/0103300.

\bibitem{Freund:2001ui}
M. Freund, P. Huber and M. Lindner,
\newblock Nucl. Phys. B615 (2001) 331, hep-ph/0105071.

\bibitem{Barger:2001yr}
V. Barger, D. Marfatia and K. Whisnant,
\newblock Phys. Rev. D65 (2002) 073023, hep-ph/0112119.

\bibitem{PDG}
Particle Data Group, K. {Hagiwara} et~al.,
\newblock {Phys. Rev.} D66 (2002) 010001+.

\bibitem{Maltoni:2003da}
M. Maltoni et~al.,
\newblock Phys. Rev. D68 (2003) 113010, hep-ph/0309130.

\bibitem{Botella:1987wy}
F.J. Botella, C.S. Lim and W.J. Marciano,
\newblock Phys. Rev. D35 (1987) 896.

\bibitem{Nicolaidis:1988fe}
A. Nicolaidis,
\newblock Phys. Lett. B200 (1988) 553.

\bibitem{Liu:1998nb}
Q.Y. Liu, S.P. Mikheyev and A.Y. Smirnov,
\newblock Phys. Lett. B440 (1998) 319, hep-ph/9803415.

\bibitem{Freund:1999vc}
M. Freund and T. Ohlsson,
\newblock Mod. Phys. Lett. A15 (2000) 867, hep-ph/9909501.

\bibitem{deGouvea:2000un}
A.~de Gouv\^ea,
Phys.\ Rev.\ D 63 (2001) 093003, hep-ph/0006157.

\bibitem{Huber:2003ak}
P. Huber and W. Winter,
\newblock Phys. Rev. D68 (2003) 037301, hep-ph/0301257.

\bibitem{Jacobson:2003wc}
M. Jacobson and T. Ohlsson,
\newblock Phys. Rev. D69 (2004) 013003, hep-ph/0305064.

\end{thebibliography}
\end{document}